\documentclass{optica-article}

\journal{opticajournal} 

\articletype{Research Article}

\usepackage{lineno}
\usepackage{pdfpages}

\begin{document}

\title{Generalized model of anisotropic thermo-optic response on thin-film lithium niobate platform}

\author{
Joonsup Shim,\authormark{1,2,$\dagger$}
Seonghun Kim,\authormark{2,$\dagger$}
Shengyuan Lu,\authormark{1}
Jiayu Yang,\authormark{1}
Seongjin Jeon,\authormark{2}
Sanghyeon Kim,\authormark{2}
Marko Lončar,\authormark{1,3,*}
and Young-Ik Sohn\authormark{2,4,*}}

\address{\authormark{1}John A. Paulson School of Engineering and Applied Sciences, Harvard University, Cambridge, Massachusetts 02138, USA\\
\authormark{2}School of Electrical Engineering, Korea Advanced Institute of Science and Technology (KAIST), 291 Daehak-Ro, Yuseong-Gu, Daejeon 34141, Republic of Korea\\
\authormark{$\dagger$}These authors contributed equally.}

\email{\authormark{3}loncar@g.harvard.edu, \authormark{4}youngik.sohn@kaist.ac.kr} 


\begin{abstract*} 
Thermo-optic (TO) control is crucial for thin-film lithium niobate (TFLN) photonic integrated circuits (PICs), offering a simple and practical method for low-frequency and DC tuning while remaining compatible with high-frequency electro-optic (EO) modulation. In x-cut TFLN, the TO response is inherently anisotropic, depending on both waveguide propagation angle and polarization due to the mode-specific overlap of the electric field with the ordinary and extraordinary refractive index axes of the crystal. Despite its significance, a systematic and quantitative analysis of this anisotropy has remained elusive. Here, we present the first generalized analytical model that describes the anisotropic TO response as a function of polarization and arbitrary waveguide orientation, and rigorously validate it through numerical simulations and experiments. This study provides foundational insight into anisotropic thermal tuning and enables new opportunities for engineering energy-efficient and scalable photonic design in next-generation TFLN PICs.
\end{abstract*}

\section{INTRODUCTION}

Thin-film lithium niobate (TFLN) has emerged as a leading platform for high-performance photonic integrated circuits (PICs), owing to its strong electro-optic (EO) and nonlinear coefficients, wide transparency window, and compatibility with heterogeneous integration \cite{Zhu.2021}. These advantages have enabled the realization of high-speed modulators  \cite{Hu.2025,Renaud.2023, Wang.20189zl, Zhang.2021wbd, Hu.2025fj9, Song.2025}, efficient frequency converters \cite{Chen.2024, Jankowski.20208wl, Wang.201827r, Xin.2025}, frequency comb generators \cite{song2025hybrid, gong2022monolithic, hu2022high}, and quantum photonic devices \cite{Warner.2025, Chapman.2025, Kim.2024} on compact, chip-scale platforms, addressing growing demands in communication, sensing, and computing. While the EO effect in TFLN is well suited for high-speed modulation, many applications such as tunable filters \cite{Ding.2022}, wavelength trimming \cite{Liu.2022}, interferometric switches \cite{Song.2023}, in-phase/quadrature (IQ) modulators \cite{Xu.2020}, and programmable photonic processors \cite{Wei.2025} require low-frequency or static tuning. In these regimes, thermo-optic (TO) phase shifters provide an effective alternative, offering a simpler and more compact solution. In particular, EO operation in TFLN under low-frequency or DC bias is often challenging due to drift and relaxation of the EO response \cite{holzgrafe2024relaxation, Powell.2024}. Although DC stability can be partially recovered at cryogenic temperatures \cite{warner2025dc}, such approaches are not ideal for most applications. As a result, TO tuning has become a widely adopted and effective approach for low-frequency and static phase control in TFLN PICs \cite{zhang2021integrated}.

Unlike isotropic materials such as silicon or silicon nitride, the TO response in lithium niobate is intrinsically anisotropic because of its uniaxial birefringence. In bulk LN, it is well known that the extraordinary refractive index exhibits a larger TO coefficient than the ordinary index ($dn_\text{e}/dT$ > $dn_\text{o}/dT$) \cite{Gayer.2008}. In integrated waveguides, however, the effective TO response becomes more complex because it depends not only on the intrinsic material anisotropy but also on the mode-specific overlap of the electric field with the ordinary and extraordinary axes. Consequently, the change in effective refractive index induced by TO tuning varies strongly with both polarization state and waveguide propagation angle. Despite the wide adoption of TFLN PICs and the extensive use of TO tuning in practical devices \cite{Chapman.2025, Kim.2024, Ding.2022, Liu.2022, Song.2023, Wei.2025, Gayer.2008, Maeder.2022}, a systematic and quantitative study of anisotropic TO modulation in x-cut TFLN waveguides has been lacking. In particular, the dependence of TO response on arbitrary waveguide orientation, polarization state, and device geometries has not been adequately addressed, which limits the systematic design of scalable TFLN PICs.

In this work, we present a generalized analytical model that captures polarization- and orientation-dependent TO effects in x-cut TFLN waveguides. The model quantitatively explains how polarization state and waveguide propagation angle relative to the crystal axes influence the TO response, and it provides practical guidelines for optimizing TO phase shifter performance. We validate the model through finite-element simulations and experimental measurements in the telecom band, thereby establishing a foundation for understanding and engineering anisotropic TO response in TFLN PICs. Furthermore, we apply the analytical framework to circular and Euler waveguide bends, demonstrating an example of design simplification where complex structures are abstracted into a single effective parameter. This study delivers quantitative design rules for TO phase shifters, enabling optimized architectures such as Mach-Zehnder interferometers, tunable resonators, and hybrid EO/TO circuits.

\section{ANALYTICAL MODEL AND SIMULATION}

To describe the anisotropic TO response in x-cut TFLN waveguides, we developed a generalized analytical model that accounts for both polarization and waveguide orientation relative to the crystal axes, and also includes the influence of buried oxide (BOX). Figure 1(a) illustrates the coordinate system and waveguide geometry considered in this study. The crystal $x$- and $y$-axes correspond to the ordinary refractive index $n_\text{o}$, while the $z$-axis corresponds to the extraordinary refractive index $n_\text{e}$. A waveguide oriented at an angle $\theta$ with respect to the $y$-axis supports fundamental TE (transverse electric) or TM (transverse magnetic) modes whose electric-field components project differently onto the ordinary and extraordinary axes. The rib waveguide geometry is defined by its width $W_\mathrm{wg}$, height $H_\mathrm{wg}$, and slab thickness $H_\mathrm{slab}$.

Figure 1(b) illustrates representative cases for propagation along the $y$-axis ($\theta = 0^\circ$) and $z$-axis ($\theta = 90^\circ$). For TE modes, the dominant transverse field component ($E_\parallel$) aligns primarily with the extraordinary axis at $\theta = 0^\circ$, while it projects mainly onto the ordinary axis at $\theta = 90^\circ$. For TM modes, $E_\parallel$ is largely aligned with the ordinary axis. These projections are the primary factors governing the anisotropic TO response in x-cut TFLN waveguides. In TM mode analysis, however, it is important to consider the contribution from the non-negligible longitudinal field components ($E_\perp$), where they introduce additional overlap with the extraordinary axis and lead to a weak but noticeable angle dependence. Simulated mode profile distributions for the cases in Fig. 1(b) are shown in Fig. S1 in Supplement 1.

The effective TO response in x-cut TFLN waveguides can be expressed as a weighted sum of the material contributions:
\begin{subequations}
    \label{eq:linear_dndTP}
    \begin{align}
        \frac{dn_\mathrm{eff}}{dT} = \Gamma_\text{o}(\theta)\frac{dn_\text{o}}{dT} + \Gamma_\text{e}(\theta)\frac{dn_\text{e}}{dT} + \Gamma_\mathrm{BOX}(\theta)\frac{dn_\mathrm{SiO_2}}{dT}, \\
        \frac{dn_\mathrm{eff}}{dP} = \Gamma_\text{o}(\theta)\frac{dn_\text{o}}{dP} + \Gamma_\text{e}(\theta)\frac{dn_\text{e}}{dP} + \Gamma_\mathrm{BOX}(\theta)\frac{dn_\mathrm{SiO_2}}{dP},
    \end{align}
\end{subequations}
where $n_\text{eff}$ is the effective refractive index of the guided mode and $n_\mathrm{SiO_2}$ is the refractive index of the buried oxide (SiO$_\text{2}$). The coefficients $\Gamma_\text{o}(\theta), \Gamma_\text{e}(\theta),$ and $\Gamma_\text{BOX}(\theta)$ represent the confinement factors in the ordinary, extraordinary, and oxide regions, respectively, defined as the optical field intensity overlapping each region per unit guided power \cite{Robinson.2008}. Although air-clad devices are assumed here, the contribution from air is neglected because both the modal overlap and TO coefficient of air are negligible \cite{Bauld.2017}. Dispersion is not included in the model; therefore, all subsequent measurements were carried out over a sufficiently narrow spectral range to minimize its influence. Assuming the temperature change ($\Delta T$) is linearly proportional to the applied heating power ($P$), Eq. (\ref{eq:linear_dndTP}a) can be rewritten in terms of $dn_i/dP$ ($i \in \{\text{e}, \text{o}, \mathrm{SiO_2}\}$), leading to Eq. (\ref{eq:linear_dndTP}b). 

As evident from Eqs. (\ref{eq:linear_dndTP}a) and (\ref{eq:linear_dndTP}b), the coefficients $dn_i/dT$ and $dn_i/dP$ are intrinsic material properties. Consequently, the key to describing anisotropic TO response lies in modeling angle-dependent confinement factors $\Gamma_i(\theta)$. To analytically model these factors, we began from the variational theorem of Maxwell's equations for the modal field profiles \cite{Kogelnik.1974}, which can be explicitly expressed for the extraordinary and ordinary contributions (see Section 1 in Supplement 1 for the full derivation):
\begin{subequations}
    \label{eq:derivative}
    \begin{align}
        \Gamma_\text{e} = \frac{\partial n_\mathrm{eff}}{\partial n_\text{e}} &= 2 c n_{\text{e}}\varepsilon_0 \frac{\int_{A_\text{LN}}|E_{z}|^2\, dA}{\int_{A_\infty}\left(\mathbf{E}\times \mathbf{H}^*+\mathbf{E}^*\times \mathbf{H}\right)\cdot\hat{r}\, dA}, \\
        \Gamma_\text{o} = \frac{\partial n_\mathrm{eff}}{\partial n_\text{o}} &= 2 c n_{\text{o}}\varepsilon_0 \frac{\int_{A_\text{LN}}\left(|E_{x}|^2 + |E_{y}|^2\right)dA}{\int_{A_\infty}\left(\mathbf{E}\times \mathbf{H}^*+\mathbf{E}^*\times \mathbf{H}\right)\cdot\hat{r}\, dA},
    \end{align}
\end{subequations}
where the numerator integrals are evaluated over the lithium niobate cross-section ($A_\mathrm{LN}$), and the denominator integral is taken over the entire cross-section ($A_\infty$). 

\begin{figure}[htbp]
    \centering\includegraphics[width=\textwidth]{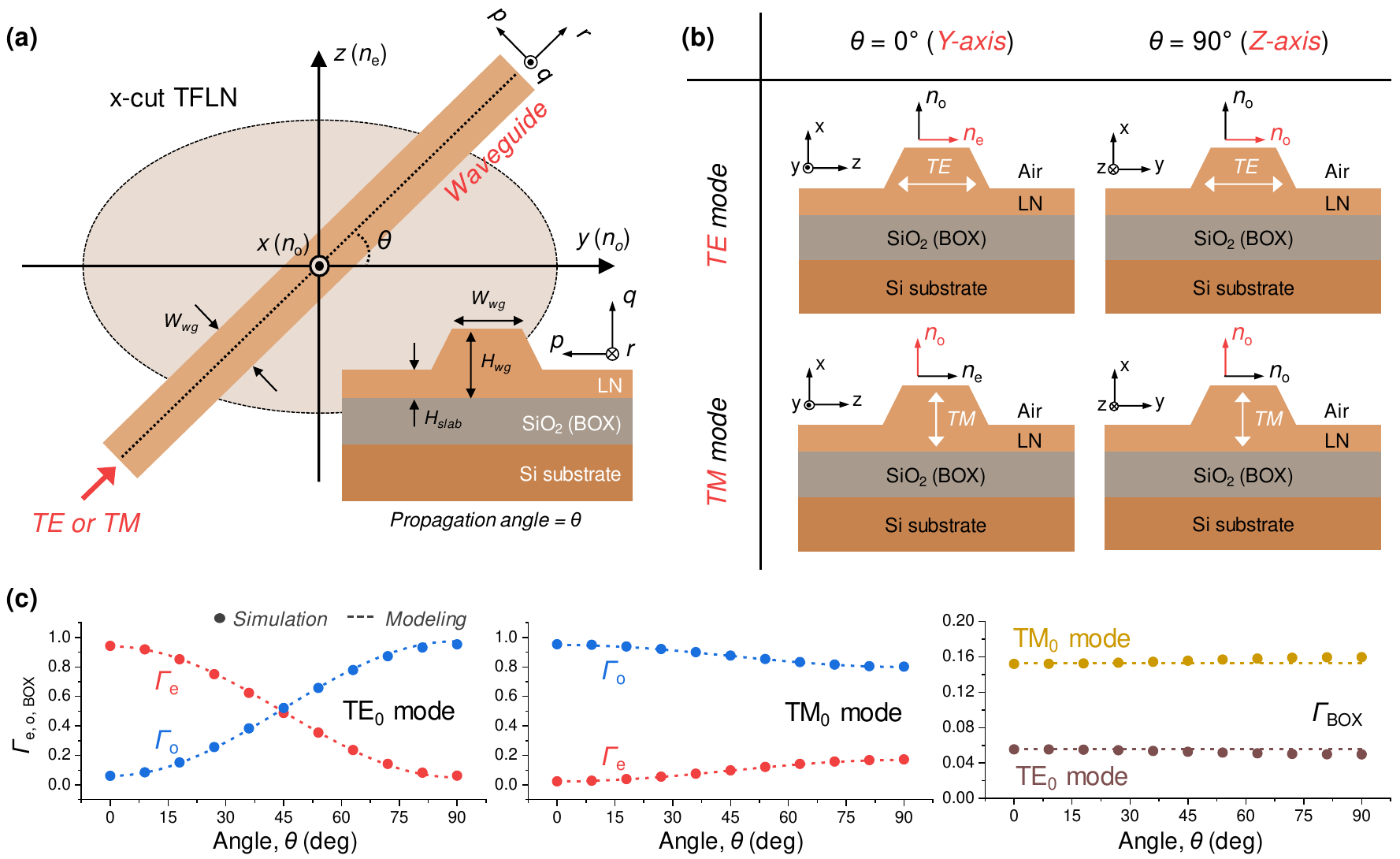}
    \caption{(a) Schematic illustration of the x-cut TFLN waveguides. Definition of the waveguide propagation angle $\theta$ with respect to the crystal axes and cross-sectional geometry of the rib waveguides are described. The local waveguide coordinate system is defined as $(p, q, r)$. (b) Representative cases of TE and TM modes propagating along the $y$-axis ($\theta = 0^\circ$) and $z$-axis ($\theta = 90^\circ$), showing how modal electric field components (transverse mode, $E_\parallel$) project onto the ordinary and extraordinary axes. (c) Comparison of angle-dependent confinement factors $\Gamma_i(\theta)$ ($i \in \{\text{e}, \text{o}, \mathrm{BOX}\}$) between our analytical model and numerical simulation results.}
    \label{fig:1}
\end{figure}

In order to describe the varying electric field projections onto the crystal axes with changing propagation angle, we introduce a local waveguide coordinate system $(p,q,r)$ in Fig. \ref{fig:1}(a). The $r$-axis is aligned with the propagation direction, while the $q$-axis is parallel to the crystal $x$-axis of the x-cut TFLN wafer. As a result, the mode propagates in the material $y$-$z$ plane, which corresponds to the $p$-$r$ plane of the local system. We then apply a coordinate transformation between the local and material systems (see Section 1 in Supplement 1). This yields the following simplified analytical expressions for $\Gamma_\text{e}$ and $\Gamma_\text{o}$:
\begin{subequations}
    \label{eq:gamma_factor_angle}
    \begin{align}
        \Gamma_\text{e} = \frac{\partial n_\mathrm{eff}}{\partial n_\mathrm{e}} &= n_\mathrm{e}\left(\Lambda_p\cos^2\theta + \Lambda_r\sin^2\theta\right), \\
        \Gamma_\text{o} = \frac{\partial n_\mathrm{eff}}{\partial n_\mathrm{o}} &= n_\mathrm{o} \left(\Lambda_p\sin^2\theta + \Lambda_q + \Lambda_r\cos^2\theta\right),
    \end{align}
\end{subequations}
where the overlap factors $\Lambda_k$ are defined as:
\begin{equation}
    \label{eq:overlap}
    \Lambda_k = 2 c \varepsilon_0 \frac{\int_{A_\text{LN}}|E_{k}|^2\, dA}{\int_{A_\infty}\left(\mathbf{E}\times \mathbf{H}^*+\mathbf{E}^*\times \mathbf{H}\right)\cdot\hat{r}\, dA}, \quad k \in \{p,q,r\}.
\end{equation}
Here, $\Lambda_k$ are assumed to be angle-independent, implying invariance of the local field components $(E_p, E_q, E_r)$ regardless of the waveguide angle $\theta$. This assumption will be validated later by evaluating the accuracy of the full model. Thus, the angle dependence of the confinement factors arises solely from the geometric rotation of the waveguide. As a result, the angle dependence of the $\Gamma_\text{e}$ and $\Gamma_\text{o}$ follows directly from the simple trigonometric relations in Eq. (\ref{eq:gamma_factor_angle}). Here, the confinement factor for the buried oxide ($\Gamma_\mathrm{BOX}$) is taken as constant.

We evaluated $\Gamma_i(\theta)$, expressed in Eq. (\ref{eq:gamma_factor_angle}), using two methods: (1) direct numerical simulation with a finite-difference eigenmode (FDE) solver (Lumerical), and (2) the analytical model. The rib waveguide geometry used in both methods is shown in Fig. 1(a), with $W_\mathrm{wg}$ = 1 {\textmu}m, $H_\mathrm{wg}$ = 600 nm, $H_\mathrm{slab}$ = 200 nm, and a symmetric sidewall angle of 60$^\circ$. In the numerical approach, we applied a small perturbation $\Delta n_i$ to each material refractive index and computed the corresponding change in $\Delta n _\mathrm{eff}$, yielding $\Gamma_i \approx \Delta n _\mathrm{eff}/\Delta n _i $. This procedure was repeated for multiple propagation angles for both TE and TM modes (see Section 2-2 in Supplement 1). In the analytical approach, $\Gamma_i(\theta)$ was calculated from Eq. (\ref{eq:gamma_factor_angle}) using overlap factors $\Lambda_k$, which were obtained once at $\theta = 0^\circ$ from numerically computed field distributions. The numerator in Eq. (\ref{eq:overlap}) was evaluated only over the lithium niobate cross-section, while the denominator integral was calculated over the entire cross-section. Figure \ref{fig:1}(c) compares the results from the two approaches, showing excellent agreement. Minor discrepancies are attributed to neglecting the weak angle dependence of $\Lambda_k$ (see Section 2-3 in Supplement 1 for more information). This confirms the validity of our analytical model and supports the assumption of angle-invariant local field profiles for x-cut TFLN waveguides.

\section{DEVICE FABRICATION AND CHARACTERIZATION}
To experimentally validate our analytical model, we designed tunable racetrack resonators on 5\% magnesium-oxide (MgO)-doped x-cut TFLN on insulator wafer, consisting of a 600 nm-thick top LN layer, a 2 {\textmu}m-thick SiO$_\text{2}$ buried oxide, and a 525 {\textmu}m-thick Si substrate. The detailed fabrication process flow is summarized in Appendix A-1. In this work, we focus on 5\% MgO-doped LN, a widely adopted material that significantly raises the photorefractive damage threshold, ensuring stable performance under high optical intensities while preserving the electro-optic and nonlinear coefficients of LN \cite{Zhu.2021}. Since MgO-doped TFLN wafers are commercially available, our study provides results directly relevant to the practical design of thermo-optic devices in TFLN PICs.

\begin{figure}[h]
    \centering\includegraphics[width=\textwidth]{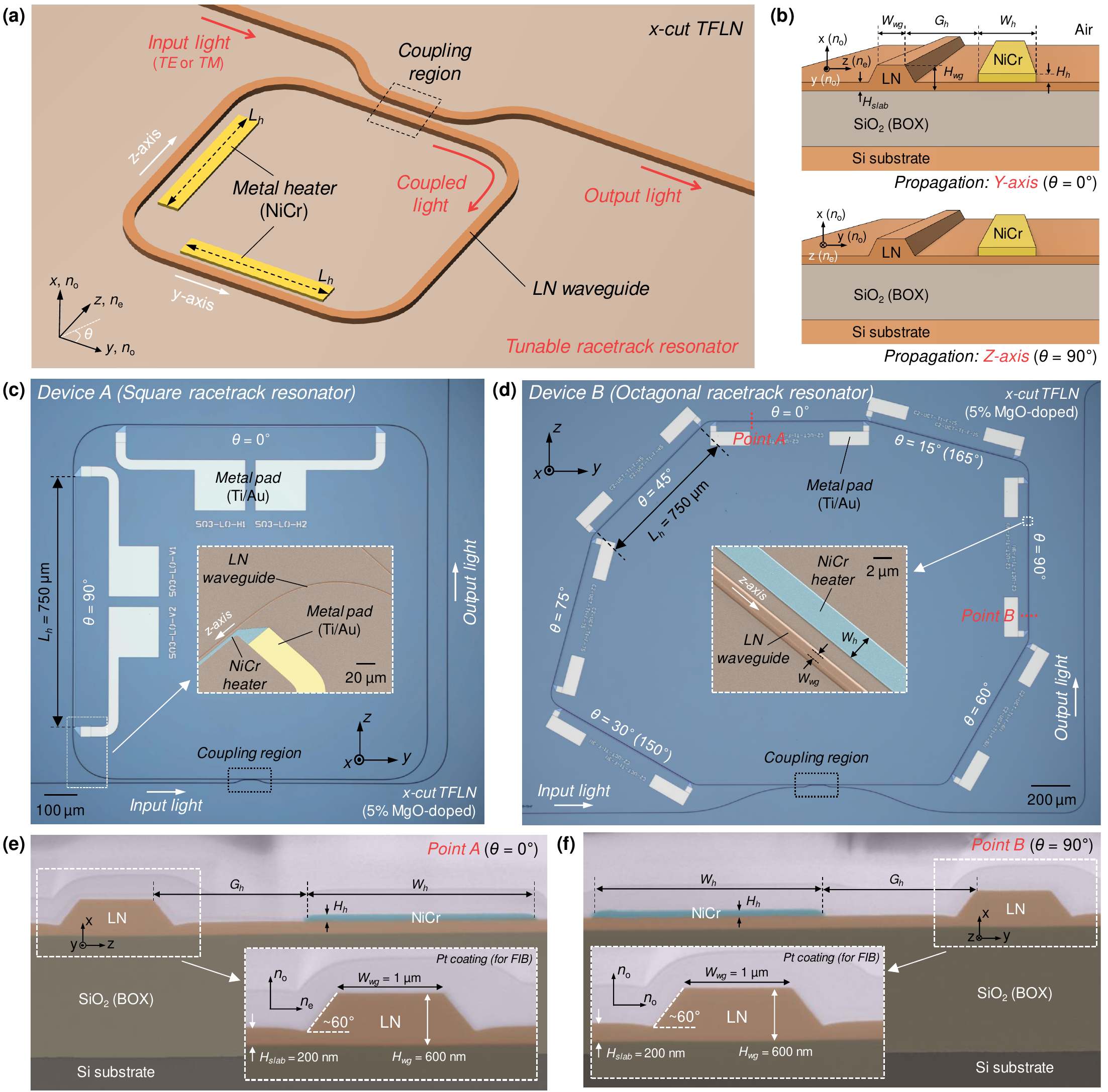}
    \caption{Device structure. (a) Schematic of the x-cut TFLN racetrack resonator with integrated metal (NiCr) heaters oriented along the $y$- and $z$-axes. (b) Cross-sectional views of the heater-waveguide geometry for propagation along the $y$-axis ($\theta = 0^\circ$) and z-axis ($\theta = 90^\circ$). Design parameters: $W_\mathrm{wg}$ = 1 {\textmu}m, $H_\mathrm{wg}$ = 600 nm, $H_\mathrm{slab}$ = 200 nm, $W_\mathrm{h}$ = 3 {\textmu}m, $L_\mathrm{h}$ = 750 {\textmu}m, $H_\mathrm{h}$ = 100 nm, and $G_\mathrm{h}$ = 2 {\textmu}m. (c) Optical microscope image of the fabricated square racetrack resonator (\textit{Device A}) with heaters aligned along $\theta = 0^\circ$ and $90^\circ$. (d) Optical microscope image of the octagonal racetrack resonator (\textit{Device B}) incorporating waveguides at multiple orientations ($\theta = 0^\circ, 15^\circ, 30^\circ, 45^\circ, 60^\circ, 75^\circ, 90^\circ$). Insets in (c) and (d) are false-colored SEM images of the magnified view around the heater-waveguide region. (e, f) Cross-sectional SEM images at \textit{Points A} ($\theta = 0^\circ$) and \textit{B} ($\theta = 90^\circ$) in (d), showing heater-waveguide spacing and layer structure (NiCr heater, LN waveguide, $\mathrm{SiO_2}$ buried oxide, and Si substrate).}
    \label{fig:2}
\end{figure}

Figure \ref{fig:2}(a) illustrates the concept of the racetrack resonator with integrated metal heaters placed along $y$- and $z$-axis ($\theta = 0^\circ$ and $90^\circ$) to explore the anisotropic TO response in x-cut TFLN. The corresponding cross-sectional geometries for propagation along $\theta = 0^\circ$ and $90^\circ$ are shown in Fig. \ref{fig:2}(b). Depending on the propagation angle of the waveguides, the optical mode projects differently onto the ordinary and extraordinary index axes of the crystal, leading to distinct TO modulation efficiencies.

Figure \ref{fig:2}(c) shows an optical microscope image of the fabricated square racetrack resonator (\textit{Device A}), which incorporates heater oriented along $\theta = 0^\circ$ and $90^\circ$. The metal heaters had identical dimensions with a width $W_\mathrm{h}$ = 3 {\textmu}m, length $L_\mathrm{h}$ = 750 {\textmu}m, height $H_\mathrm{h}$ = 100 nm, and heater-waveguide gap $G_\mathrm{h}$ = 2 {\textmu}m. The gap $G_\mathrm{h}$  was chosen to mitigate additional propagation loss from metal-induced absorption (see Section 3 in Supplement 1). Nickel-chromium (NiCr) was selected as the heater material due to its high electrical resistivity and low temperature coefficient of resistance, which enable stable and reliable TO tuning \cite{Ji.2025}. Electrical contact pads (Ti/Au) were deposited as a separate layer to ensure that current flowed only through the NiCr heaters. The measured current-voltage (\textit{I-V}) characteristics confirmed ohmic behavior (see Section 4 in Supplement 1). To extend the study beyond two orientations, we also fabricated an octagonal racetrack resonator (\textit{Device B}), shown in Fig. \ref{fig:2}(d), which integrates heaters along multiple propagation angles ($\theta = 0^\circ, 15^\circ, 30^\circ, 45^\circ, 60^\circ, 75^\circ, 90^\circ$). This configuration enables a systematic investigation of the orientation dependence of the TO response across a broader range of crystal axes. Detailed design processes of the racetrack resonator are described in Section 5 in Supplement 1.

Finally, Fig. \ref{fig:2}(e) and Fig. \ref{fig:2}(f) present cross-sectional SEM images at $\theta = 0^\circ$ (\textit{Point A}) and $\theta = 90^\circ$ (\textit{Point B}), respectively. These confirm the fabricated heater-waveguide geometries, including heater-waveguide spacing and etched thicknesses of waveguide, in agreement with the design parameters. Together, \textit{Devices A} and \textit{B} allow a comprehensive characterization of anisotropic TO response in x-cut TFLN waveguides.

To directly visualize the temperature distribution in our device under electrical bias, we employed thermoreflectance microscopy (TRM), which enables thermal mapping with high spatial resolution in electronic and photonic devices \cite{Baek.2023, Lim.2024, Kim.20257wj, Jeong.2024, Shim.2022, Shim.20229mk}. In this work, TRM was used to quantitatively analyze the thermal behavior of the fabricated racetrack resonator under bias applied to NiCr heaters oriented along the $y$- and $z$-axes. The TRM setup and detailed measurement results are described in Section 6 in Supplement 1.

\section{MEASUREMENTS AND RESULTS}

With the fabricated devices characterized, we next investigated the TO response of the racetrack resonators under electrical bias. Optical transmission spectra were measured by coupling the fundamental TE or TM mode from a tunable laser into the waveguide and tracking resonance peak shifts as a function of applied heater power. The detailed measurement setup is described in Appendix A-2. Using the octagonal racetrack resonator (\textit{Device B}), we systematically explored the dependence of TO tuning on both waveguide orientation and polarization.

\begin{figure}[h]
    \centering\includegraphics[width=\textwidth]{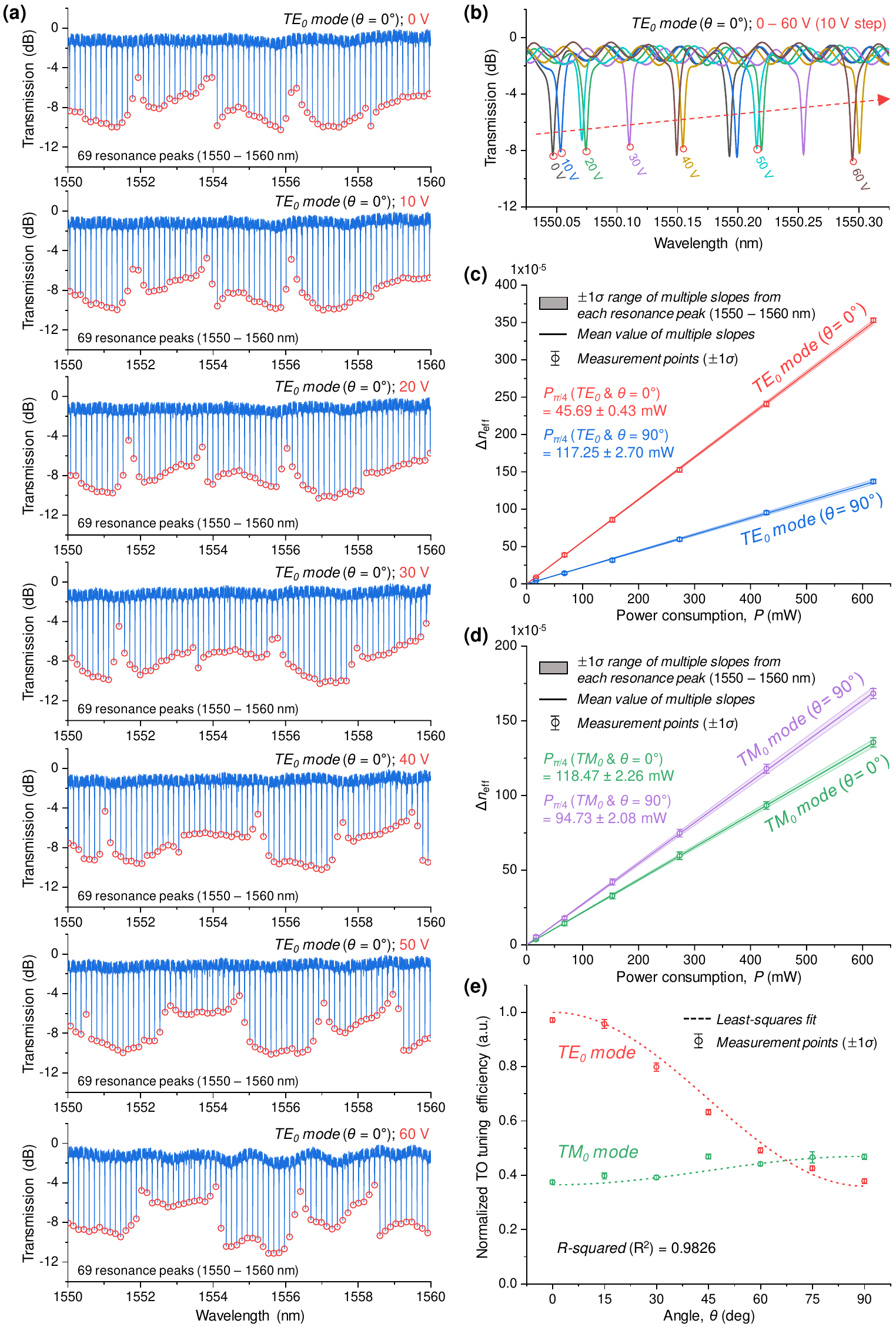}
    \caption{Experimental results of TO tuning in the octagonal racetrack resonator (\textit{Device B}). (a) Transmission spectra with TE-polarized light (1550-1560 nm) under applied heater voltages from 0 to 60 V in 10 V steps (heater at $\theta = 0^\circ$). Resonance peak shifts are tracked to extract changes in effective refractive index ($\Delta n _\text{eff}$). (b) Enlarged view of a representative resonance peak shifts under bias. (c, d) Extracted $\Delta n _\text{eff}$ versus power consumption ($P$) for (c) TE and (d) TM modes. Slopes yield the TO tuning efficiency, and the power required for a phase shift $\Delta\phi=\pi/4\ (P_{\pi/4})$ is indicated. (e) Normalized TO tuning efficiency as a function of polarization and propagation angle. Experimental data are fitted using a least-squares method with confinement factors from the analytical model, showing strong agreement ($R^2 = 0.9826$).}
    \label{fig:3}
\end{figure}

To minimize the influence of dispersion, all measurements were performed within a narrow wavelength range of 1550-1560 nm. Figure \ref{fig:3}(a) shows representative transmission spectra for the TE mode at $\theta = 0^\circ$ under incrementally increased heater voltage.
As illustrated in Fig. \ref{fig:3}(b), resonance peaks were tracked with increasing bias to extract the resonance wavelength shifts ($\Delta\lambda$). The relationship between $\Delta\lambda$ and effective index change ($\Delta n _\text{eff}$) is expressed as
\begin{equation}
    \Delta\phi = 2\pi\frac{\Delta\lambda}{\text{FSR}} = \frac{2\pi\Delta n_\text{eff}}{\lambda}L_h,
\end{equation}
where $\Delta\phi$ is the induced phase shift, FSR is the free spectral range of the resonator, and $L_h$ is the heater length. Since the $\Delta n_\text{eff}$ calculation needs the value of wavelength-dependent FSR, we extracted it from the measurement and applied it to obtain the value of $n_\text{eff}$. See Section 8 in Supplement 1 for the details of FSR extraction. From these procedures, $\Delta n_\text{eff}$ can be expressed as a function of heater power consumption ($P$), as shown in Figs. \ref{fig:3}(c) and \ref{fig:3}(d). Here, the slope of these plots ($\Delta n_\text{eff}/\Delta P$) corresponds directly to the TO tuning efficiency. Resonance peak shifts within the 1550-1560 nm range were all analyzed individually with varying $P$, and the resulting slope values were averaged. Figures \ref{fig:3}(c) and \ref{fig:3}(d) show the mean values together with $\pm1\sigma$ deviation. Note that the deviations are mostly caused by the uncertainty in the resonance fitting and tracking, rather than fundamental FSR change due to the dispersion. From these slopes, we further calculated the power required for a $\pi/4$ phase shift ($P_{\pi/4}$), providing a practical benchmark for device performance. Complete measurement data for all orientation-polarization combinations are summarized in Section 9 in Supplement 1.

The TO tuning efficiencies obtained from the slopes ($\Delta n_\text{eff}/\Delta P$) are equivalent to the left-hand side of Eq. (\ref{eq:linear_dndTP}b), assuming linear dependence of $\Delta n_\text{eff}$ on the applied power $P$. As described in Eq. (\ref{eq:linear_dndTP}b), the angular dependence is governed by the confinement factors $\Gamma_i(\theta)$ ($i \in \{\text{e}, \text{o}, \text{BOX}\}$), which were calculated using our analytical model derived in Eq. (\ref{eq:gamma_factor_angle}). Using these $\Gamma_i(\theta)$ values together with 14 experimental data points ($\theta = 0^\circ, 15^\circ, 30^\circ, 45^\circ, 60^\circ, 75^\circ, 90^\circ$ for TE and TM modes) and three unknown fitting parameters ($\frac{dn_\text{o}}{dP}$, $\frac{dn_\text{e}}{dP}$, and $\frac{dn_\mathrm{SiO_2}}{dP}$) in Eq. (\ref{eq:linear_dndTP}b), the system of linear equations is overdetermined. We therefore applied a least-squares fitting approach to quantitatively assess the agreement between our model and the measurements.

Figure \ref{fig:3}(e) shows the normalized TO tuning efficiencies along with the corresponding fits. The experimental data exhibit strong agreement with the least-squares fit ($R^2 = 0.9826$), confirming both the validity and predictive capability of our analysis. Here, a clear anisotropic trend is observed: TE modes show strong dependence on propagation angle, with the tuning efficiency maximized near $\theta = 0^\circ$ (aligned with the extraordinary axis) and minimized near $\theta = 90^\circ$ (aligned with the ordinary axis). In contrast, TM modes are primarily aligned with the ordinary axis and thus exhibit much weaker angular dependence. A slight orientation dependence nonetheless remains for TM modes, arising from non-negligible longitudinal field components ($E_\perp$) that partially overlap with the extraordinary axis, which depends on the propagation angle $\theta$. This observation is consistent with the predictions of our analytical model shown in Fig. \ref{fig:1}(c) and highlights the fundamental role of modal field projections in governing anisotropic TO response. Overall, these results establish, for the first time, a comprehensive and quantitative understanding of the anisotropic TO response in x-cut TFLN waveguides, linking the dependence on polarization and propagation orientation to the underlying crystal anisotropy.

\section{DISCUSSION AND OUTLOOK}

In this work, we proposed a generalized analytical model for anisotropic TO response in x-cut TFLN waveguides and validated it by experimental studies. In addition to establishing the anisotropic TO model, further insights can be obtained by building upon it. An important aspect is the influence of waveguide geometry. Variations in waveguide width and slab thickness could modify mode confinement and thus affect the absolute TO efficiency. Simulations presented in Section 10 in Supplement 1 confirm that such geometrical variations do not significantly influence the TO tuning efficiency for each polarization and propagation direction, reinforcing the generality of our study.

Moreover, many practical photonic circuits incorporate bent waveguides such as circular or Euler bends \cite{vogelbacher2019analysis}. To address this, we further extended our analytical model to bent geometries with arbitrary bend parameters.
The average TO coefficient for a phase shifter placed on a non-straight waveguide is defined as the integral of the local TO coefficient over the path length:
\begin{equation}\label{eq:ave_dndT}
    \left(\frac{dn_\mathrm{eff}}{dT}\right)_\mathrm{ave} = \frac{1}{L_h}\int  \frac{dn_\mathrm{eff}}{dT}dl,
\end{equation}
where the integral is taken along the waveguide path $l$, and $L_h$ is the total heated length.
Our aim is to define an equivalent propagation angle, $\theta_\text{eq}$, for a bent waveguide such that a straight waveguide oriented at $\theta_\text{eq}$ exhibits the identical average TO response over length $L_h$, as illustrated in Fig. \ref{fig:4}(a). A complete derivation is provided in Section 11 in Supplement 1.
Figure \ref{fig:4}(e) presents the equivalent propagation angles for three representative geometries: a $\pi/2$-bend ($\alpha = 90^\circ$), a $\pi$-bend starting at $0^\circ$ ($\alpha = 180^\circ$, symmetry plane along the $y$-axis), and a $\pi$-bend starting at $90^\circ$ ($\alpha = 180^\circ$, symmetry plane along the $z$-axis) shown in Figs. \ref{fig:4}(b)-(d), respectively. Here, $\alpha$ denotes the total bend angle, and $p$ is the partial Euler fraction. For the $\pi$-bend symmetric about the $y$-axis, $\theta_\text{eq}$ decreases with increasing $p$, whereas for the $\pi$-bend symmetric about the $z$-axis, $\theta_\text{eq}$ increases as $p$ increases. This behavior arises because a larger Euler fraction $p$ causes the waveguide to spend more of its length near the initial and final tangent directions. This extended model enables predictive design of bent waveguides with tailored TO responses, allowing bending geometries to be selected or engineered based on a quantitative understanding of their TO behavior.

The cladding environment is another key consideration. Here, our analysis and measurements focused on air-clad devices, whereas SiO$_\text{2}$ top cladding is also widely used. The addition of SiO$_\text{2}$ modifies the optical mode overlap of the guided field with each material layer, thereby altering the effective TO response. To quantify this effect, we performed simulations of anisotropic TO tuning efficiency for SiO$_\text{2}$-clad waveguides (see Section 12 in Supplement 1). These results indicate that while the absolute efficiency changes slightly, particularly for TM modes, the polarization- and orientation-dependent trends revealed in this study remain consistent.

Although our model attributes refractive index changes primarily to the TO effect, other mechanisms such as fabrication imperfections (e.g., alignment errors), thermally induced photoelastic effect, and pyroelectric-induced electro-optic contributions \cite{Zhu.2021, Pan.2023, Ren.2025} may also play secondary roles, introducing slight deviations from the trends observed in Fig. \ref{fig:3}(e). However, the strong agreement between our model and the measurements suggests that these effects are minor under our operating conditions, and that the TO effect dominates the observed response.

The implications of this study extend further to nonlinear and quantum photonics. TO control plays a critical role in thin-film periodically poled lithium niobate (TF-PPLN) devices, where wavelength tuning \cite{Liu.2022}, recovery of distorted quasi-phase-matching (QPM) spectra \cite{Li.2024}, and active phase tuning between cascaded PPLN devices \cite{Kim.2024} are required. Therefore, our anisotropy-aware model is not only useful for classical devices, such as modulators and tunable filters, but also for nonlinear and quantum photonic circuits.

While we primarily focused on the analysis of the anisotropic behavior of TO phase tuning rather than maximizing absolute efficiency, our model can provide baseline information to design advanced heater architectures, such as cladding-embedded heaters, thermally isolated trenches \cite{Liu.2022}, or folded waveguide layouts \cite{Ji.2025}, to further enhance performance. Our polarization- and orientation-aware model can be readily applied to such configurations and accelerate integration in large-scale PICs.
Looking forward, the analytical model developed here is not restricted to MgO-doped TFLN and is directly applicable to congruent TFLN platforms. More broadly, it can be readily extended to other anisotropic material platforms such as thin-film lithium tantalate (TFLT) \cite{Wang.2024, Hulyal.2025, Powell.2024, Powell.2025, Shelton.2025} and barium titanate (BTO) \cite{Wen.2024, Kim.2025}, as well as to operation across different spectral regions, including the visible and mid-infrared.

It is worth noting that one straightforward method to improve the inherently lower TO tuning efficiency of TM modes compared to TE modes is to exploit mode conversion through a tapered waveguide along the $y$-axis in x-cut TFLN. In this scheme, the fundamental TM$_0$ mode is converted into the first-order TE$_1$ mode, which exhibits stronger overlap with the extraordinary axis and thus higher TO efficiency. After thermal tuning, the light can be converted back from TE$_1$ to TM$_0$ using an inverted taper. Simulation results provided in Section 13 in Supplement 1 confirm the feasibility of this approach, showing that mode conversion offers a practical strategy to overcome the intrinsically weaker TO response of TM modes and expand the design space for anisotropic PICs.

In conclusion, this work presents the first systematic and quantitative model for anisotropic TO control in x-cut (MgO-doped) TFLN waveguides. By developing an analytical model validated through simulations and experiments, we demonstrated how the TO response is governed by polarization, propagation direction, and device geometry. Based on the validated model, we further extended the model to predict the TO response of bent waveguides. Beyond validating the physics of the anisotropic TO response, this study provides a practical design tool that can be directly applied to photonic circuit designs based on anisotropic material platforms, ensuring predictable and energy-efficient operation. Taken together, these results provide clear design guidelines and a general principle for anisotropy-aware thermal control, opening new opportunities for energy-efficient, scalable, and multifunctional photonic architectures on the x-cut TFLN platform for both classical and quantum applications.

\section*{APPENDIX A: METHODS}
\section*{A-1. DEVICE FABRICATION}

The fabrication process began with a commercial 5\% magnesium-oxide (MgO)-doped x-cut thin-film lithium niobate (TFLN) on insulator wafer (NANOLN), which consists of a 600 nm-thick top LN, a 2  {\textmu}m-thick SiO$_\text{2}$ buried oxide layer, and a 525 {\textmu}m-thick Si substrate. Electron-beam lithography with a negative-tone resist (ma-N 2405) was first performed to define passive structures, including rib waveguides and racetrack resonators. Afterwards, Ar$^+$-based reactive ion etching was employed to achieve a 400 nm etch depth, followed by wet cleaning with hot SC-1 and piranha solutions to remove re-deposition and residual resist. Subsequently, we conducted a second electron-beam lithography step to pattern the metal heaters and then deposited 100 nm-thick nickel-chromium (NiCr) through electron-beam evaporation, followed by lift-off. Finally, contact electrodes of Ti/Au (15/200 nm) were formed using a combination of photolithography, electron-beam evaporation, and lift-off process.

\section*{A-2. DETAILS OF MEASUREMENTS}

For optical measurements, a tunable continuous wave (CW) laser (Santec TSL-570) was employed as the input light source. The polarization state was carefully adjusted using a fiber polarization controller and a polarizing beam-splitting (PBS) cube, allowing selection of either TE or TM polarization. The polarized light was coupled into the waveguide facet of the fabricated sample using a lensed fiber, and the output light from the opposite facet was collected with a lensed fiber and detected by a photoreceiver (Newport 1811-FC). To operate the metal heaters, electrical probes connected to a benchtop DC power supply (Keithley 2200-72-1) were used to apply bias voltages.

\bibliography{heater_paper_cite}

\newpage

\begin{figure}[h]
    \centering\includegraphics[width=\textwidth]{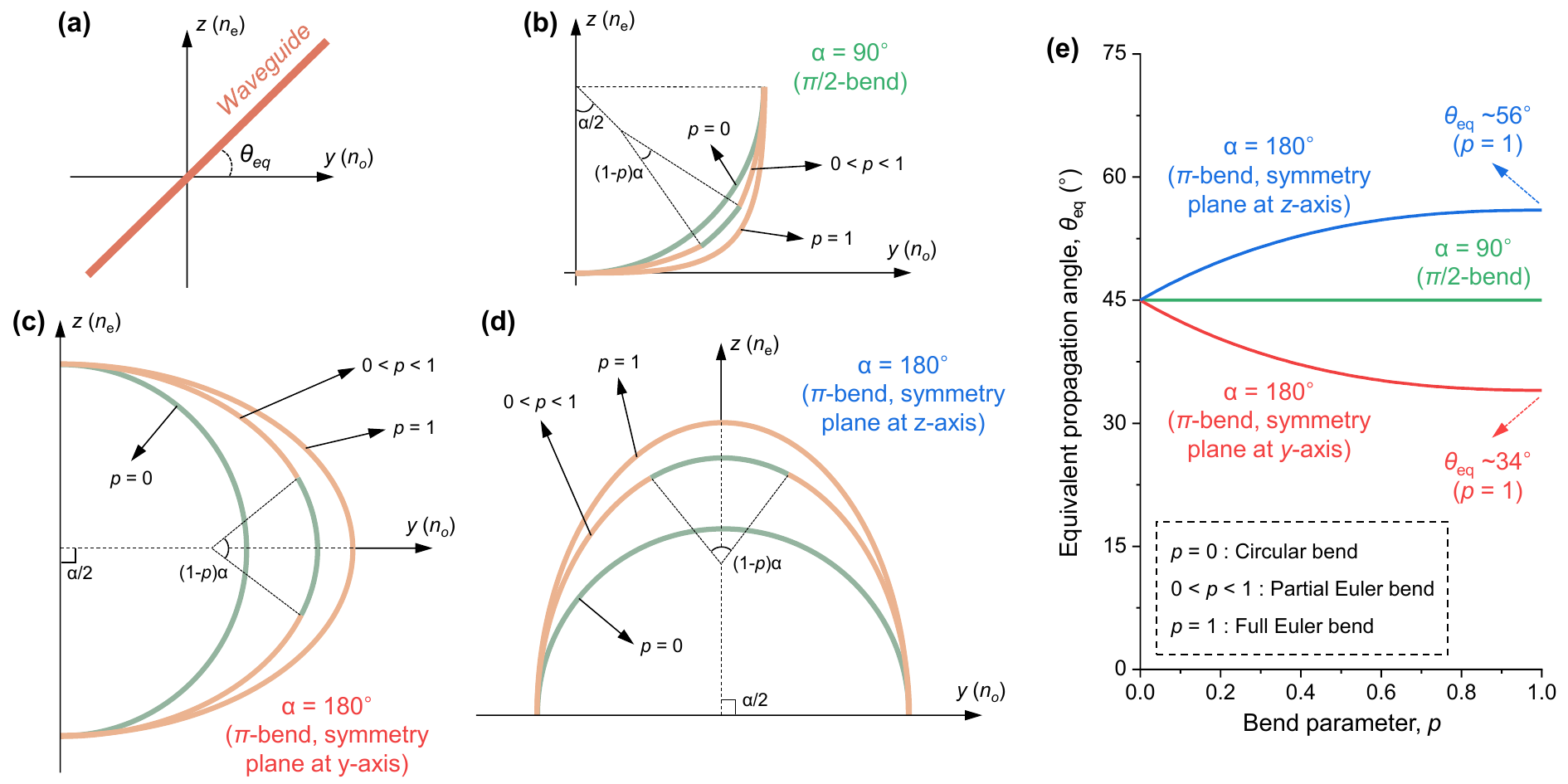}
    \caption{
    Equivalent propagation angle for TO analysis in bent waveguides. (a) Conceptual illustration of the equivalent propagation angle, $\theta_\text{eq}$. (b-d) Schematics of the three representative partial Euler bend geometries analyzed: (b) a $\pi/2$-bend, (c) a $\pi$-bend starting with a tangent angle of 0$^\circ$, and (d) a $\pi$-bend starting with a tangent angle of 90$^\circ$. (e) Calculated equivalent propagation angle $\theta_\text{eq}$ as a function of the partial Euler fraction parameter $p$ for three representative geometries.}
    \label{fig:4}
\end{figure}

\clearpage


\section*{Supplementary material}

# Generalized model of anisotropic thermo-optic response on thin-film lithium niobate platform: supplemental document

## 1. DERIVATION DETAILS OF ANALYTICAL MODELING

We start from the variational theorem of Maxwell's equation [1], which relates a perturbation in the propagation constant ($\beta$) to perturbations in frequency ($\omega$) and permittivity ($\varepsilon$):

$$\delta\beta = \frac{\int (\mathbf{E} \cdot \delta(\omega\varepsilon)\mathbf{E}^* + \mathbf{H} \cdot \delta(\omega\mu)\mathbf{H}^*) \, dA}{\int (\mathbf{E} \times \mathbf{H}^* + \mathbf{E}^* \times \mathbf{H}) \cdot \hat{r} \, dA}, \tag{S1}$$

where the integrals are taken over the waveguide cross-section ($dA$) and $\mathbf{r}$ is unit vector in direction of propagation. The vectors $\mathbf{E}$ and $\mathbf{H}$ are the complex amplitude of the electric and magnetic modal field of the waveguide. Here, non-magnetic materials and constant frequency are assumed:

$$\delta\mu = 0, \quad \delta\omega = 0. \tag{S2}$$

Assuming a diagonal permittivity tensor,

$$\varepsilon = \begin{bmatrix} \varepsilon_x & 0 & 0 \\ 0 & \varepsilon_y & 0 \\ 0 & 0 & \varepsilon_z \end{bmatrix}, \tag{S3}$$

the variation of $\beta$ with respect to a perturbation in $\varepsilon_i$ becomes

$$\frac{\partial \beta}{\partial \varepsilon_i} = \omega \frac{\int_{A_\varepsilon} |E_i|^2 \, dA}{\int_{A_\infty} (\mathbf{E} \times \mathbf{H}^* + \mathbf{E}^* \times \mathbf{H}) \cdot \hat{r} \, dA}, \quad i \in \{x, y, z\}, \tag{S4}$$

where the numerator is integrated over the region with perturbed refractive index ($A_\varepsilon$), and the denominator over the entire cross section ($A_\infty$).

Using the relation $\beta = 2\pi n_{\text{eff}}/\lambda$ and $\varepsilon = \varepsilon_0 n_i^2$, where $n_{\text{eff}}$ is the effective refractive index, $\lambda$ is the vacuum wavelength, and $n_i$ is the material refractive index along a crystal axis ($i \in \{x, y, z\}$), Eq. (S4) can be rewritten as:

$$\frac{\partial n_{\text{eff}}}{\partial n_i} = 2c n_i \varepsilon_0 \frac{\int_{A_\varepsilon} |E_i|^2 \, dA}{\int_{A_\infty} (\mathbf{E} \times \mathbf{H}^* + \mathbf{E}^* \times \mathbf{H}) \cdot \hat{r} \, dA}. \tag{S5}$$

Here, we note that the right-hand side of Eq. (S5) corresponds to the definition of confinement factors, $\Gamma_i$ [2]. For uniaxial crystals such as lithium niobate, these can be expressed explicitly as the extraordinary and ordinary contributions:

$$\Gamma_e = \frac{\partial n_{\text{eff}}}{\partial n_{\text{e}}} = 2c n_{\text{e}} \varepsilon_0 \frac{\int_{A_\varepsilon} |E_z|^2 \, dA}{\int_{A_\infty} (\mathbf{E} \times \mathbf{H}^* + \mathbf{E}^* \times \mathbf{H}) \cdot \hat{r} \, dA}, \tag{S6a}$$

$$\Gamma_o = \frac{\partial n_{\text{eff}}}{\partial n_{\text{o}}} = 2c n_{\text{o}} \varepsilon_0 \frac{\int_{A_\varepsilon} (|E_x|^2 + |E_y|^2) \, dA}{\int_{A_\infty} (\mathbf{E} \times \mathbf{H}^* + \mathbf{E}^* \times \mathbf{H}) \cdot \hat{r} \, dA}, \tag{S6b}$$

where $n_{\text{e}}$ and $n_{\text{o}}$ are the extraordinary and ordinary material refractive indices, respectively.

To derive the Eq. (3) of the main article, we introduce the local waveguide coordinate system $(p, q, r)$ (defined in Fig. 1(a) of the main article) and apply a rotation between the local system and

the crystal coordinate system $(x, y, z)$. The transformation between two systems can be expressed with the propagation angle $\theta$:

$$\begin{bmatrix} E_x \\ E_y \\ E_z \end{bmatrix} = \begin{bmatrix} 0 & 1 & 0 \\ -\sin\theta & 0 & \cos\theta \\ \cos\theta & 0 & \sin\theta \end{bmatrix} \begin{bmatrix} E_p \\ E_q \\ E_r \end{bmatrix}. \tag{S7}$$

Substituting Eq. (S7) into Eq. (S6) yields the simplified expressions for $\Gamma_e$ and $\Gamma_o$ used in the Eq. (3) of the main article.

Similarly, the confinement factor corresponding to the buried oxide (BOX) region is

$$\frac{\partial n_{\text{eff}}}{\partial n_{\text{SiO}_2}} = 2cn_{\text{SiO}_2}\varepsilon_0 \frac{\int_{A_{\text{SiO}_2}} |\mathbf{E}|^2 \, dA}{\int_{A_\infty} (\mathbf{E} \times \mathbf{H}^* + \mathbf{E}^* \times \mathbf{H}) \cdot \hat{r} \, dA}, \tag{S8}$$

where the numerator is integrated over the region with material $SiO_2$.

## 2. SIMULATION RESULTS OF ANALYTICAL MODELING

### 2-1. Mode profiles of representative geometries

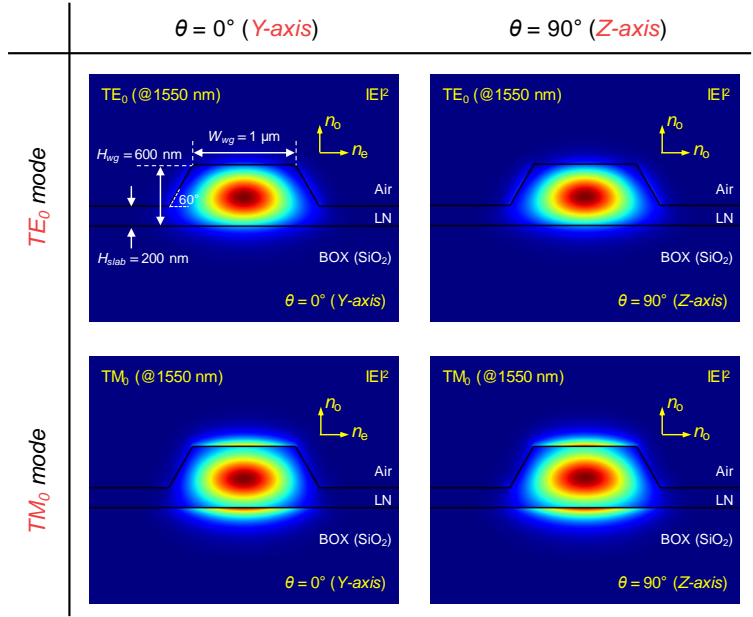

**Fig. S1.** Simulated electric field profiles $|\mathbf{E}|^2$ for the fundamental $TE_0$ and $TM_0$ modes at a wavelength of 1550 nm. The fields were obtained using a Finite Difference Eigenmode (FDE) solver (Lumerical) for waveguide propagation along the crystal $y$-axis ($\theta = 0°$) and $z$-axis ($\theta = 90°$). The black outline marks the waveguide geometry.

Fig. S1 shows the simulated field intensity distributions of the $TE_0$ and $TM_0$ modes for the representative propagation directions $\theta = 0°$ and $\theta = 90°$. These mode profiles form the basis for evaluating the confinement factors in the analytical model and highlight the distinct electric-field distributions associated with the ordinary and extraordinary axes of the x-cut TFLN platform.

### 2-2. Confinement factors via index perturbation

This section describes the numerical procedure to calculate the confinement factors ($\Gamma_i$, $i \in \{e, o, SiO_2\}$) with a finite-difference eigenmode (FDE) solver, which are later used to validate the analytical model.

The confinement factor is defined as $\Gamma_i = \partial n_{\text{eff}}/\partial n_i \approx \Delta n_{\text{eff}}/\Delta n_i$, determined by calculating the change in effective index ($\Delta n_{\text{eff}}$) due to a small perturbation in a material refractive index

$\Delta n_i$. To evaluate this, $\Delta n_{\text{eff}}$ was computed for 11 perturbation values of $\Delta n_i$, from 0 to $5 \times 10^{-4}$ in steps of $5 \times 10^{-5}$, where $\Delta n_i = 0$ corresponds to the unperturbed waveguide.

This process was repeated for both $TE_0$ and $TM_0$ modes at 11 propagation angles from $0°$ to $90°$ in $9°$ increments. The resulting linear dependence of $\Delta n_{\text{eff}}$ on $\Delta n_i$ is shown in Fig. S2. A least-squares linear fit was applied to extract the slope, which yields the confinement factor $\Gamma_i$. For all linear regressions, the coefficient of determination ($R^2$) exceeded 0.9999, ensuring that the index perturbation remained well within the linear regime. Each extracted slope corresponds to one confinement factor for a specific mode and propagation angle, and these values are from the data points plotted in Fig. 1(c) of the main article.

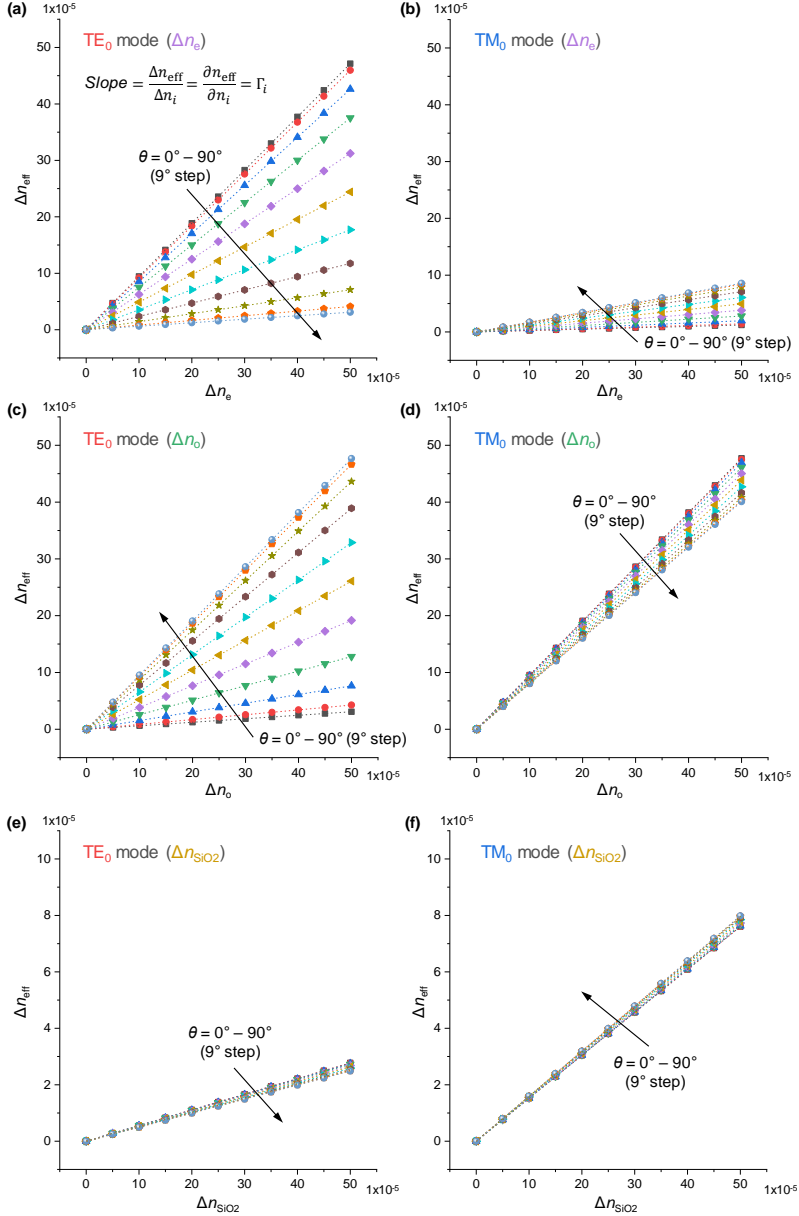

**Fig. S2.** Linear relationship between the change in effective index ($\Delta n_{\text{eff}}$) and material index perturbations for the $TE_0$ and $TM_0$ modes. Each subplots corresponds to perturbations applied to a different refractive index: $\Delta n_{\text{e}}$, $\Delta n_{\text{o}}$, and $\Delta n_{\text{SiO}_2}$.

3

## 2-3. Comparison of efficiency predictions from model and simulation

As shown in Fig. 1(c) of the main article, small discrepancies exist between the confinement factors predicted by our analytical model and those obtained from numerical simulation. Here, we evaluate whether these differences meaningfully affect the predicted thermo-optic (TO) responses.

To perform this comparison, we used both sets of confinement factors (analytical and simulated) to calculate the TO tuning efficiencies using the same experimentally fitted parameters to the model. Figure S3(a) plots the resulting efficiency curves, which are nearly indistinguishable for both $TE_0$ and $TM_0$ modes. The percentage difference between the predictions is plotted in Fig. S3(b), revealing maximum deviations below 1.5% for the $TE_0$ mode and below 2% for the $TM_0$ mode across all propagation angles.

These results confirm that the simplifying assumptions used in our analytical model (e.g., angle-invariant modal overlap factors) introduce only negligible error in the predicted TO response. Consequently, the analytical model remains highly accurate while offering substantial computational simplicity compared to full numerical simulations.

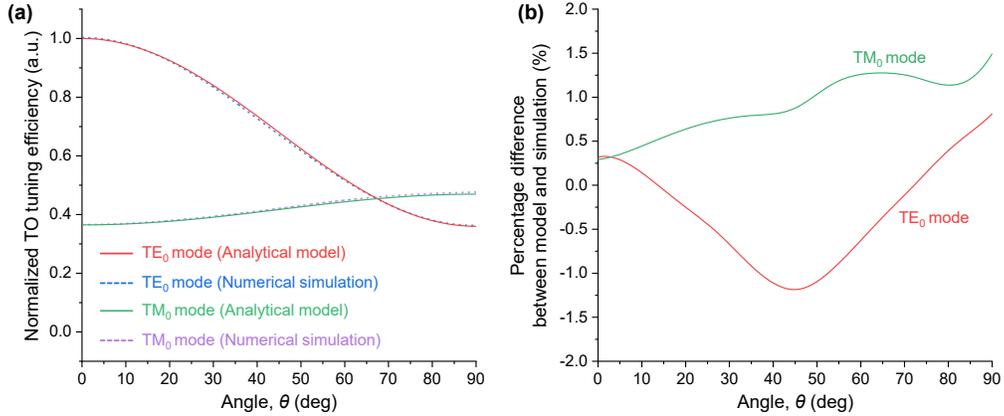

**Fig. S3.** Comparison of predicted TO tuning efficiencies using confinement factors from the analytical model and numerical simulation. (a) Predicted efficiency as a function of propagation angle for $TE_0$ and $TM_0$ modes. (b) Percent difference between the two predictions, with maximum deviations below 1.5% and 2% for $TE_0$ and $TM_0$ modes, respectively.

## 3. METAL-INDUCED ABSORPTION LOSS

The heater-to-waveguide gap, $G_h$ (defined in Fig. 2(b) of the main article), was chosen to ensure that the excess propagation loss induced by the metallic heater remains negligible compared to the intrinsic loss of the waveguide. To determine an appropriate value, the excess loss was simulated as a function of $G_h$ using an eigenmode solver. The results are presented in Fig. S4.

Based on this analysis, a gap of $G_h$ = 2 μm was selected. At this distance, the simulated excess losses for the fundamental TE and TM modes are approximately 0.10 dB/m and 0.25 dB/m, respectively. These values are significantly smaller than the estimated intrinsic propagation loss of the waveguide, confirming that the heater introduces minimal optical absorption as intended.

## 4. CURRENT-VOLTAGE (*I-V*) CHARACTERISTICS OF METAL HEATER

For DC current-voltage (*I-V*) characterization of the NiCr heater, a semiconductor parameter analyzer (Agilent 4156C) in a probe station with source measure units (SMUs) shown in Fig. S5(a) was used to conduct a voltage sweep from -60 V to +60 V with a 200-mV step, while measuring the current with high resolution and accuracy. As shown in Fig. S5(b), the heater exhibits a nearly-linear *I-V* response, demonstrating Ohmic behavior between the NiCr metal strip and the electrode pads (Ti/Au). This confirms stable electrical performance and reliable current injection for subsequent thermo-optic measurements.

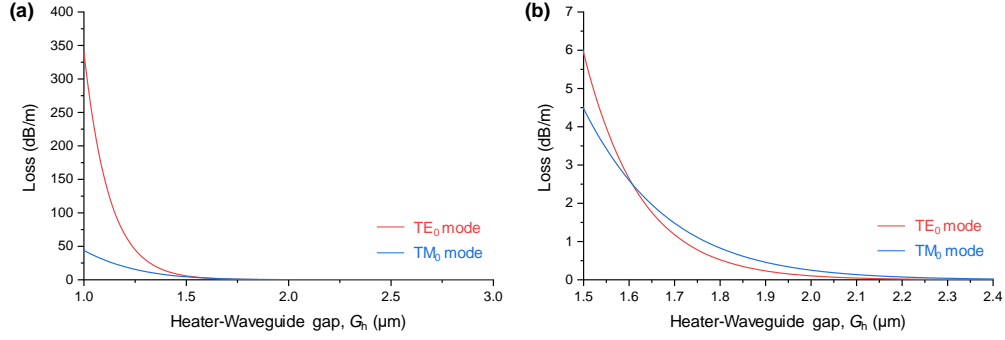

**Fig. S4.** Simulated excess propagation loss induced by the metallic heater as a function of the heater-to-waveguide gap ($G_h$). (a) Calculated loss for the fundamental TE and TM modes over a gap range of 1.0 µm to 3.0 µm. (b) Zoomed-in view of the critical region from 1.5 µm to 2.4 µm.

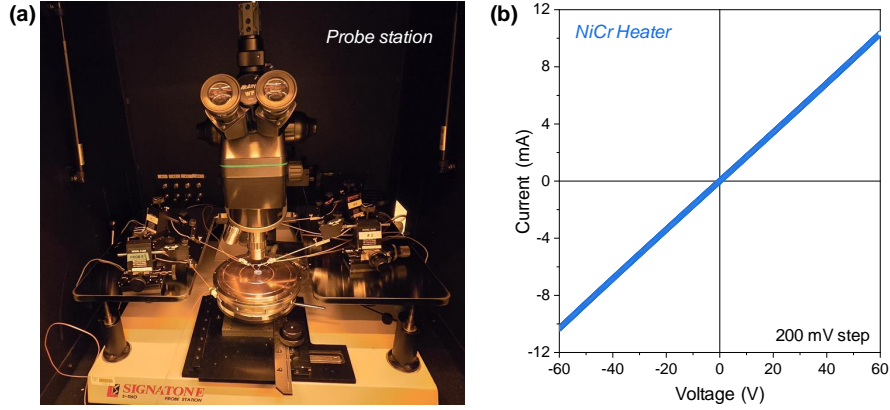

**Fig. S5.** (a) Photograph of the probe station used for *I-V* characterization. (b) Measured *I-V* curve of the NiCr heater from -60 V to +60 V with a 200-mV step.

## 5. DESIGN PROCESS OF THE RACETRACK RESONATOR

### 5-1. Coupler design

The directional couplers in the racetrack resonators were designed to operate near the critical coupling condition. In this regime, where the power coupled from the bus waveguide matches the total round-trip loss of the resonator, the extinction ratio is maximized, enabling sharper resonance features and reducing uncertainty in the extracted TO tuning parameters.

The target coupling values were determined from the measured propagation losses of waveguides fabricated using an identical process: 0.59 dB/cm for the $TE_0$ mode and 0.94 dB/cm for the $TM_0$ mode. Given the octagonal racetrack resonator's round-trip length of 7.1 mm, the corresponding round-trip power losses (and thus target coupling values) are 9.1% for $TE_0$ and 14.2% for $TM_0$.

The coupler consists of a straight coupling section placed between input and output Bézier S-bends, as illustrated in Fig. S6. The coupler gap $G_c$ was fixed at 800 nm, ans the S-bends were defined by a transverse shift of $\Delta y = 10$ µm over a longitudinal distance of $\Delta x = 50$ µm. A hybrid simulation strategy was used to model the coupler. The S-parameters of the S-bend transitions were simulated using 3D FDTD, while the propagation constants of the even and odd supermodes in the straight coupling region were obtained from a FDE solver. These components were combined in Lumerical INTERCONNECT to compute the total power coupling as a function of the coupling length $L_c$. The resulting coupling curves are plotted in Fig. S7. From these analyses, the optimal coupling lengths required to achieve critical coupling were determined to be 41.5 µm for $TE_0$ and 61.8 µm for $TM_0$.

Because a single geometry was needed to characterize both polarizations, a compromise length

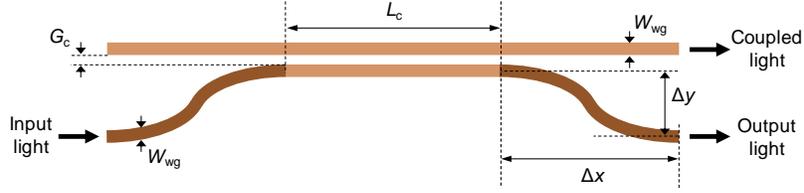

**Fig. S6.** Schematic of the directional coupler geometry. The device consists of a straight coupling region of length $L_c$ and gap $G_c$, positioned between two symmetric Bézier S-bends. The S-bends are defined by a transverse displacement $\Delta y$ over a longitudinal distance $\Delta x$. For the final design, the fixed parameters are $G_c = 800$ nm, $\Delta y = 10$ µm, and $\Delta x = 50$ µm.

of 51.6 µm was selected for fabrication. At this geometry, the simulated coupling values are 12.39% for $TE_0$ (slightly over-coupled) and 10.89% for $TM_0$ (slightly under-coupled).

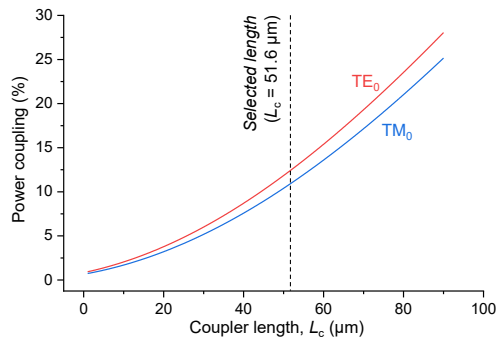

**Fig. S7.** Simulated power coupling versus straight coupling length ($L_c$) for the $TE_0$ and $TM_0$ modes. The vertical dashed line marks the chosen compromise length of 51.6 µm. At this length, the simulated couplings are 12.4% for $TE_0$ (over-coupled) and 10.9% for $TM_0$ (under-coupled).

**5-2. Mode hybridization**

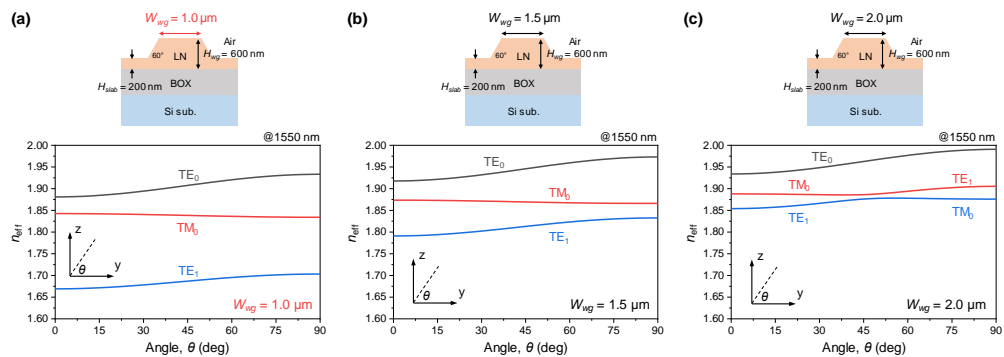

**Fig. S8.** Simulated effective indices of the first three eigenmodes as a function of propagation angle ($\theta$) for three different waveguide widths: (a) $W_{wg} = 1.0$ µm, (b) $W_{wg} = 1.5$ µm, and (c) $W_{wg} = 2.0$ µm. In (c), a pronounced mode anti-crossing is observed between the $TM_0$ and $TE_1$ modes.

The waveguide width is a critical design parameter for ensuring clean and unambiguous spectra in racetrack resonators. To extract TO tuning characteristics accurately, it is essential to suppress unintended coupling between the fundamental mode and higher-order modes.

In anisotropic waveguides, the effective indices of the fundamental and higher-order modes vary with the propagation angle. At certain angles, two modes may become nearly degenerate, leading to mode hybridization (anti-crossing). If the bent sections of a resonator force light to propagate through these angle-dependent hybridization regions, unwanted intermodal coupling can occur, causing excess loss, resonance broadening, or distorted spectral features.

To avoid this issue, we simulated the effective indices of the guided modes as a function of propagation angle for different waveguide widths. As shown in Fig. S8, a waveguide width of 2.0 µm exhibits a clear anti-crossing between the $TM_0$ and $TE_1$ modes. In contrast, a width of 1.0 µm maintains sufficient separation between the fundamental and higher-order modes across the entire angular range.

Based on this analysis, a waveguide width of 1.0 µm was chosen for fabrication. This ensures robust single-mode operation throughout the resonator path, preventing hybridization-induced coupling and enabling stable, high-quality resonance spectra for TO tuning characterization.

## 6. THERMOREFLECTANCE MICROSCOPY (TRM) CHARACTERIZATION

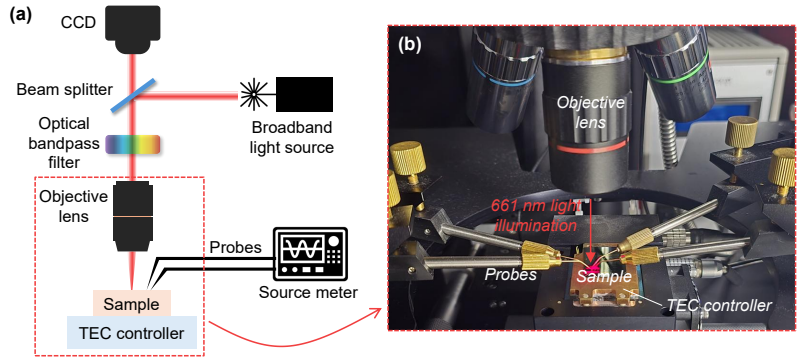

**Fig. S9.** Experimental setup for thermoreflectance microscopy (TRM). (a) Schematic of the TRM measurement system. (b) Photograph of the TRM setup with the PIC sample mounted on the stage.

Thermoreflectance microscopy (TRM) is a non-contact, non-invasive optical technique that detects small changes in surface reflectivity induced by local temperature variations. As shown in Fig. S9, TRM system (Nanoscope Systems, TRM250) was employed to characterize the temperature distribution of the fabricated racetrack resonator under electrical bias. To convert reflectivity change into absolute temperature, the thermoreflectance coefficient ($\kappa$) was calibrated prior to imaging. The calibration was performed using a charge-coupled device (CCD) camera operating at 30 fps for 60 seconds under 661 nm illumination from a broadband light source combined with a bandpass filter. During this process, the background temperature was swept using a thermoelectric cooler (TEC), enabling measurement of the reflectivity change of the TFLN waveguide core region. This yielded a thermoreflectance coefficient of $\kappa = 1.952 \times 10^{-4}$ K$^{-1}$. Following calibration, a 1 Hz square-wave voltage signal was applied to the NiCr heater to induce periodic heating. TRM images were finally extracted from the CCD camera through a Fourier-domain filtering method.

Figure S10 presents TRM surface temperature maps of the square racetrack resonator (*Device A*) under a 60 V bias (∼619.6 mW electrical power) applied to the NiCr heaters oriented at $\theta = 0°$ and $\theta = 90°$. In both cases, a pronounced temperature rise is observed around the adjacent waveguides, with the maximum waveguide temperature ($T_{\max}$) reaching ∼160°C, confirming efficient electrical-to-thermal conversion. Importantly, the overall temperature rise and spatial distribution are nearly identical between the two heater orientations. This result supports the assumption in our analytical model that the anisotropy of TO response originates primarily from the optical mode overlap with the anisotropic LN indices, rather than from orientation-dependent heating efficiency. These characterizations therefore validate that both *y*- and *z*-axis heaters, as well as arbitrarily oriented heaters, provide comparable thermal environments during the optical experiments.

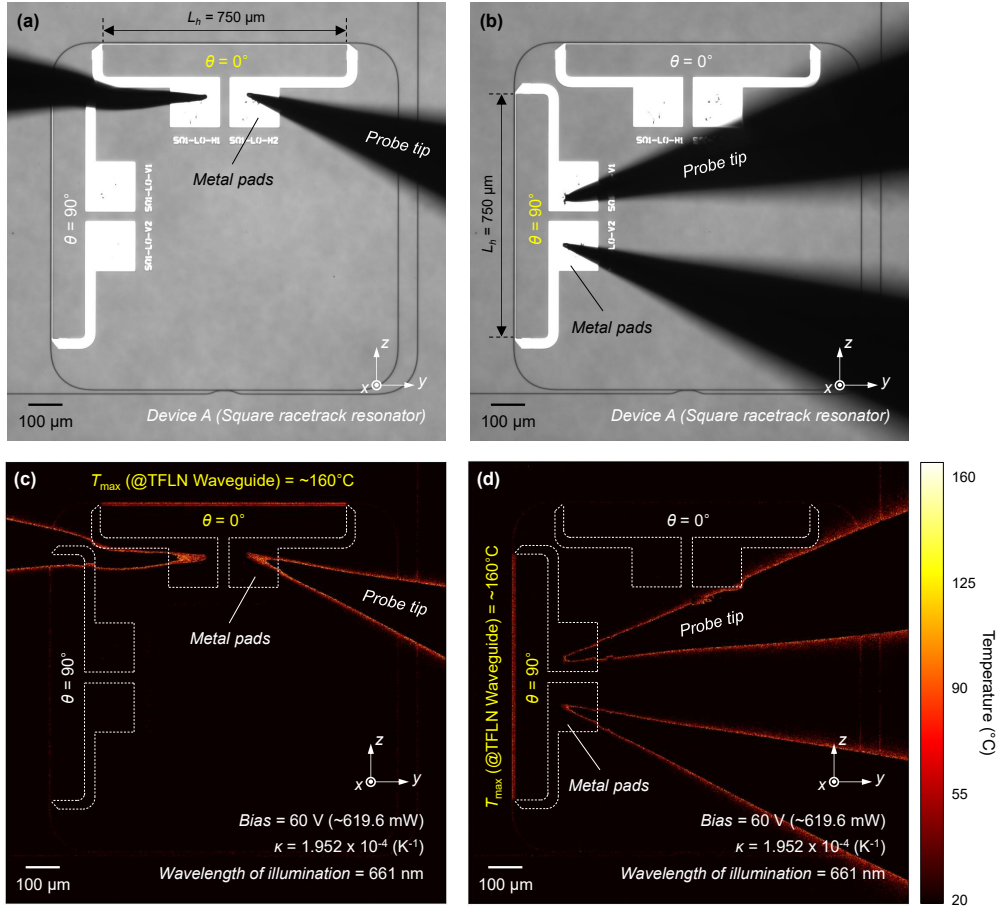

**Fig. S10.** Thermal characterization of the square racetrack resonator (*Device A*) with thermoreflectance microscopy (TRM). (a, b) CCD images under heater operation at (a) $\theta = 0°$ and (b) $\theta = 90°$. (c, d) Corresponding TRM surface temperature maps for (c) $\theta = 0°$ and (d) $\theta = 90°$. The thermoreflectance coefficient ($\kappa = 1.952 \times 10^{-4}$ K$^{-1}$) was calibrated in the TFLN waveguide region under 661 nm illumination, so the color bar is valid only within the waveguide. With an applied bias of 60 V (∼619.6 mW electrical power), the maximum waveguide temperature ($T_{\max}$) reached ∼160°C for both heater orientations. The ambient background temperature was maintained at 20°C.

## 7. OPTICAL Q-FACTOR ANALYSIS

To evaluate the optical performance of the fabricated racetrack resonators, we extracted the loaded Q-factors from the measured transmission spectra. Representative resonances with high extinction ratios were selected for both the TE$_0$ and TM$_0$ modes, as shown in Fig. S11.

For the TE$_0$ mode, a resonance at $\lambda = 1597.02$ nm was fitted using a Lorentzian profile, resulting in a loaded Q-factor ($Q_L$) of $2.29 \times 10^5$ and an extinction ratio (ER) of 13.38 dB. Similarly, the TM$_0$ mode exhibited a resonance at $\lambda = 1591.92$ nm with a loaded Q-factor of $2.17 \times 10^5$ with an ER of 11.06 dB.

The high ER values indicate that, near these wavelengths, the coupling condition approaches critical coupling, where the coupling loss and intrinsic round-trip loss are comparable. In this regime, the intrinsic Q-factor ($Q_i$), dominated by internal propagation loss, can be approximated as $Q_i \approx 2Q_L$. Using this relation, the intrinsic Q-factors are estimated to be approximately $4.58 \times 10^5$ for the TE$_0$ and $4.34 \times 10^5$ for the TM$_0$.

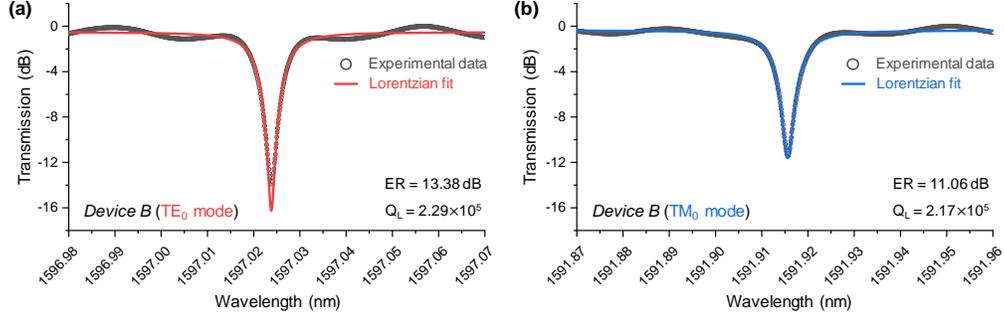

**Fig. S11.** Representative transmission spectra and Lorentzian fits for the racetrack resonator. (a) $TE_0$ mode resonance at 1597.02 nm with a loaded Q-factor ($Q_L$) of $2.29 \times 10^5$ and extinction ratio (ER) of 13.38 dB. (b) $TM_0$ mode resonance at 1591.92 nm with $Q_L$ of $2.17 \times 10^5$ and ER of 11.06 dB.

## 8. FREE SPECTRAL RANGE (FSR) EXTRACTION

To extract the heater-induced phase shift ($\Delta\phi$) from a measured resonance wavelength shift ($\Delta\lambda$), an accurate estimation of the free spectral range (FSR) at the specific resonance wavelength is required. For this purpose, the wavelength-dependent FSR was characterized for each device configuration (i.e., each polarization and propagation angle) using its zero-bias transmission spectrum.

Wavelength-dependent FSR values were obtained at each resonance peak by averaging the wavelength spacing to its two nearest neighbors. Figure S12 shows example FSR values for the $TE_0$ and $TM_0$ modes of the octagonal racetrack resonator at $\theta = 0°$. These discrete FSR values were then fitted using a linear model, FSR($\lambda$):

$$\text{FSR}(\lambda) = a + b\lambda, \tag{S9}$$

yielding parameters (a, b) of (-0.164, $1.99 \times 10^{-4}$) for $TE_0$ and (-0.187, $2.08 \times 10^{-4}$) for $TM_0$. Because the resonance spectra were measured over a narrow 10 nm wavelength span, this linear approximation is sufficiently accurate. Once the wavelength-dependent FSR is determined, the phase shift induced by electrical heating is obtained from the measured resonance shift. A resonance initially located at $\lambda_0$ under zero bias shifts by an amount $\Delta\lambda$ when a voltage is applied. The corresponding phase shift is then calculated using Eq. (5) of the main article:

$$\Delta\phi = 2\pi \frac{\Delta\lambda}{\text{FSR}(\lambda_0)}. \tag{S10}$$

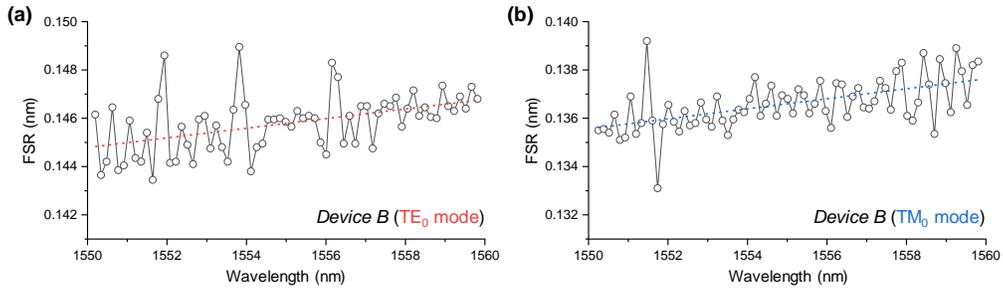

**Fig. S12.** Example FSR fitting results for the (a) $TE_0$ and (b) $TM_0$ modes. Discrete data points correspond to local FSR values extracted by averaging the wavelength spacing to the nearest adjacent resonances. Dashed lines show linear fits using Eq. (S9).

## 9. MEASUREMENT RESULTS ACROSS ALL CONDITIONS

This section provides the complete experimental dataset of effective index change ($\Delta n_{\text{eff}}$) as a function of applied heating power for all combinations of polarization ($TE_0$ and $TM_0$) and

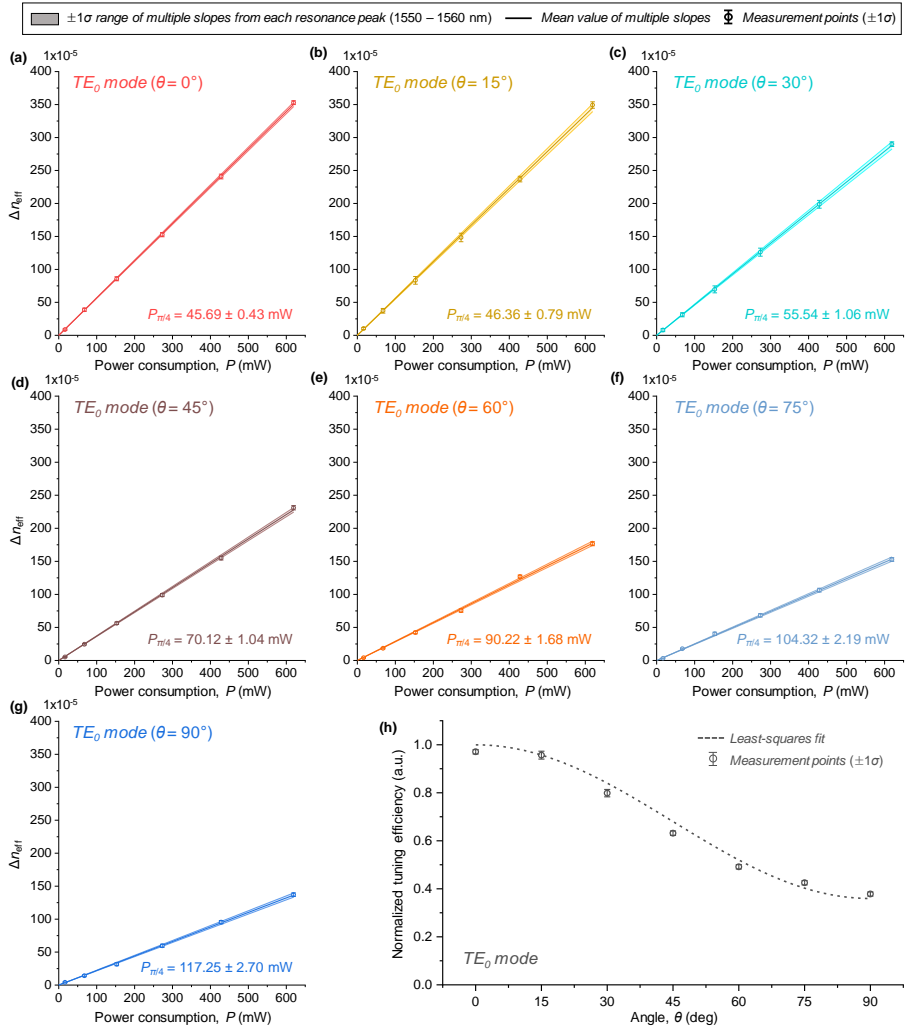

**Fig. S13.** Full experimental data for the TE$_0$ mode. (a-g) Measured effective index change ($\Delta n_{\text{eff}}$) versus heating power ($P$) for propagation angles from $\theta = 0°$ to $90°$ in $15°$ increments. The mean value of each linear fit corresponds to the TO efficiency $dn_{\text{eff}}/dP$. (h) Normalized TO tuning efficiency as a function of propagation angle, summarizing the TE$_0$ results.

propagation angle ($\theta = 0°$–$90°$). These results, presented in Figs. S13 and S14, include the full set of raw data underlying the representative examples shown in Figs. 4(c) and 4(d) of the main article. Each subplot shows the individual resonance tracking and linear fit used to extract the TO tuning efficiency ($dn_{\text{eff}}/dP$), while the final subplots (TE: Fig. S13(h); TM: Fig. S14(h)) summarize the angular dependence of the normalized tuning efficiency for each polarization.

## 10. WAVEGUIDE-GEOMETRY DEPENDENCE OF THERMO-OPTIC RESPONSE

To confirm that the strong anisotropy in TO response is an intrinsic property of x-cut TFLN platform, rather than an artifact of a particular waveguide geometry, we evaluated how the TO efficiency varies across different cross-sectional designs.

We simulated the TO efficiency for the fundamental TE and TM modes propagating along the $y$-axis ($\theta = 0°$) and $z$-axis ($\theta = 90°$), varying the waveguide width ($W_{\text{wg}}$) and slab thickness ($H_{\text{slab}}$). For each geometry, the TO efficiencies were calculated using our analytical model, incorporating the fitted material parameters from the main article (Fig. 4(e)) and the simulated mode overlap factors for each geometry ($\Lambda_k$). For meaningful comparison, all efficiencies were normalized to the TE$_0$ efficiency at $\theta = 0°$ for the nominal geometry ($W_{\text{wg}} = 1$ µm, $H_{\text{slab}} = 200$ nm).

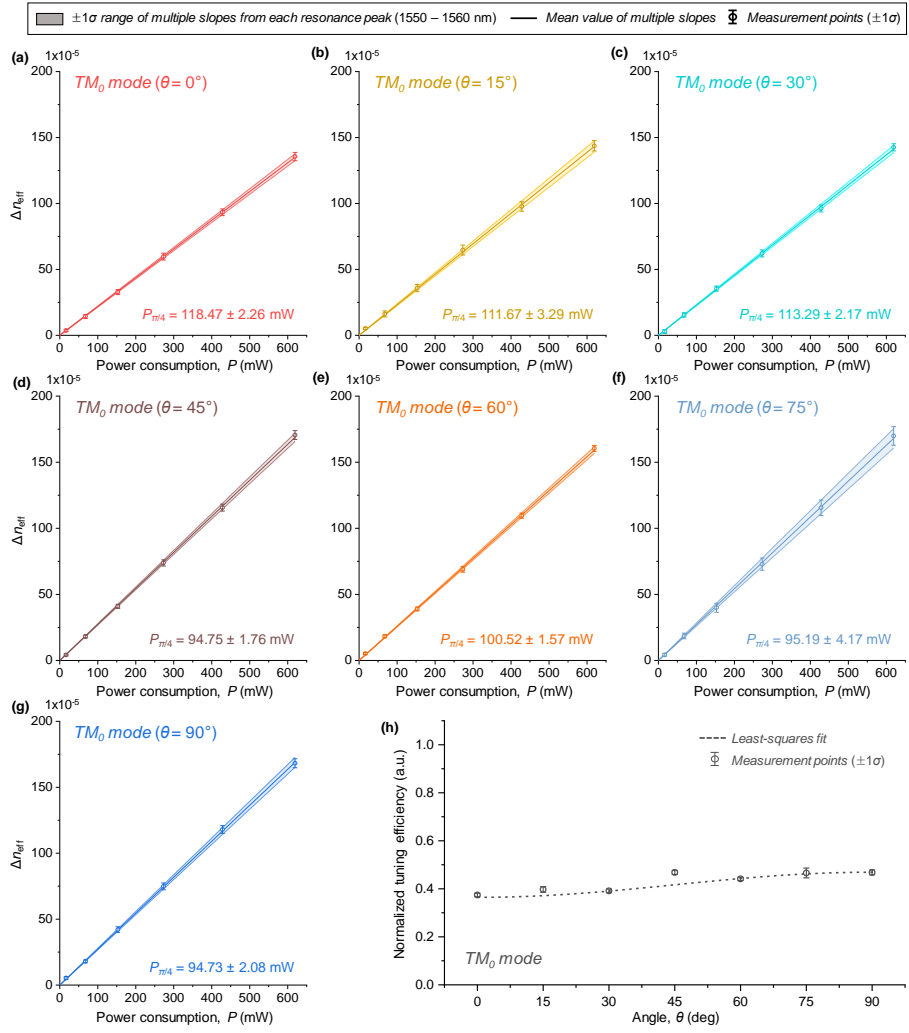

**Fig. S14.** Full experimental data for the TM$_0$ mode. (a-g) Measured effective index change ($\Delta n_{\text{eff}}$) versus heating power ($P$) for propagation angles from $\theta = 0°$ to $90°$ in $15°$ increments. The mean value of each linear fit corresponds to the TO efficiency $dn_{\text{eff}}/dP$. (h) Normalized TO tuning efficiency as a function of propagation angle, summarizing the TM$_0$ results.

As shown in Fig. S15, the anisotropic behavior of the TO response persists across all simulated geometries. Although the absolute efficiencies vary slightly with waveguide width and slab thickness, these geometric effects are small compared to the inherent differences between polarizations and propagation directions. This confirms that the anisotropic TO behavior reported in this work is robust and holds over a broad range of practical device geometries.

## 11. THERMO-OPTIC RESPONSE IN BENT WAVEGUIDES

We start from Eq. (6) in the main article:

$$\left(\frac{dn_{\text{eff}}}{dT}\right)_{\text{ave}} = \frac{1}{L_h} \int \frac{dn_{\text{eff}}}{dT} dl, \tag{S11}$$

11

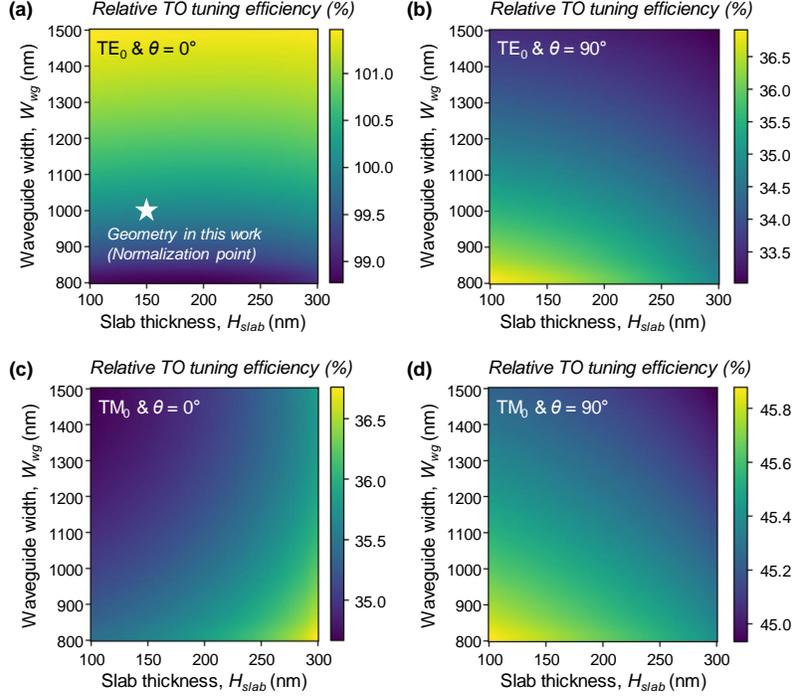

**Fig. S15.** Waveguide-geometry dependence of the relative TO efficiency. The efficiency is plotted as a function of waveguide width ($W_{\text{wg}}$) and slab thickness ($H_{\text{slab}}$), normalized to the $TE_0$ value at $\theta = 0°$ in the nominal geometry ($W_{\text{wg}} = 1$ µm, $H_{\text{slab}} = 200$ nm). The analysis includes the fundamental (a, b) $TE_0$ and (c, d) $TM_0$ modes propagating along the crystal $y$-axis ($\theta = 0°$) and $z$-axis ($\theta = 90°$).

where the integral is taken along the waveguide path $l$, and $L_h$ is the total heated length. Substituting the analytical model (Eqs. (1) and (3) in the main article) into Eq. (S11) yields:

$$\left(\frac{dn_{\text{eff}}}{dT}\right)_{\text{ave}} = n_o \frac{dn_o}{dT}\left(\Lambda_p \Theta_s + \Lambda_q + \Lambda_r \Theta_c\right) \\ + n_e \frac{dn_e}{dT}\left(\Lambda_p \Theta_c + \Lambda_r \Theta_s\right) + \Gamma_{\text{BOX}} \frac{dn_{\text{SiO}_2}}{dT}, \quad \text{(S12)}$$

where $\Theta_s$ and $\Theta_c$ are the length-averaged trigonometric factors:

$$\Theta_s = \frac{1}{L_h}\int_0^{L_h} \sin^2 \theta(l)\, dl, \quad \text{(S13a)}$$

$$\Theta_c = \frac{1}{L_h}\int_0^{L_h} \cos^2 \theta(l)\, dl. \quad \text{(S13b)}$$

By definition, $\Theta_s + \Theta_c = 1$. We therefore define the equivalent propagation angle $\theta_{\text{eq}}$:

$$\sin^2 \theta_{\text{eq}} \equiv \Theta_s, \quad \cos^2 \theta_{\text{eq}} \equiv \Theta_c. \quad \text{(S14)}$$

We evaluate Eqs. S13 and S14 explicitly for a partial Euler bend, which consists of two Euler-transition sections surrounding a central circular arc. The propagation angle $\theta(l)$ for a bend starting from $\theta(0) = 0$ is given by [3]:

$$\theta(l) = \begin{cases} \dfrac{\alpha p}{2L_e^2} l^2 & 0 \leq l \leq L_e \\ \dfrac{\alpha p}{2} + \dfrac{\alpha p}{L_e}(l - L_e) & L_e \leq l \leq L_e + L_c \\ \alpha - \dfrac{\alpha p}{2L_e^2}(w - L_e)^2 & L_e + L_c \leq l \leq L_h \end{cases} \quad \text{(S15)}$$

12

where $\alpha$ is the total bend angle and $p$ is the partial Euler fraction. The Euler-section length ($L_e$) and the circular-section length ($L_c$) are defined as:

$$L_e = \frac{pL_h}{1+p}, \quad L_c = \frac{(1-p)}{1+p}L_h, \quad w = z - (L_e + L_c). \tag{S16}$$

Evaluating Eq. (S13) for this path and substituting into Eq. (S14) yields:

$$\theta_{eq} = \frac{1}{2}\arccos(X), \tag{S17}$$

with

$$X = \frac{p}{1+p}\frac{C(\sqrt{\alpha p}) + \cos(2\alpha)C(\sqrt{\alpha p}) + \sin(2\alpha)S(\sqrt{\alpha p})}{\sqrt{\alpha p}}$$
$$+ \frac{1}{2\alpha(1+p)}\left[\sin(2\alpha - \alpha p) - \sin(\alpha p)\right]. \tag{S18}$$

Here, $C(x)$ and $S(x)$ are the Fresnel integrals:

$$C(x) = \int_0^x \cos(t^2)\, dt, \tag{S19a}$$

$$S(x) = \int_0^x \sin(t^2)\, dt. \tag{S19b}$$

The general expression for $\theta_{eq}$ can be simplified for several specific cases. For a purely circular bend ($p = 0$), the equivalent propagation angle is given in closed-form:

$$\theta_{eq} = \frac{1}{2}\arccos\left(\frac{\sin(2\alpha)}{2\alpha}\right). \tag{S20}$$

For full Euler bends with $p = 1$, it becomes:

$$\theta_{eq} = \frac{1}{2}\arccos\left(\frac{C(\sqrt{\alpha}) + \cos(2\alpha)C(\sqrt{\alpha}) + \sin(2\alpha)S(\sqrt{\alpha})}{2\sqrt{\alpha}}\right). \tag{S21}$$

For any partial Euler bend with $\alpha = 90°$, the equivalent angle becomes $\theta_{eq} = 45°$ independent of $p$. For bends with $\alpha = 180°$, the equivalent angle becomes:

$$\theta_{eq} = \frac{1}{2}\arccos\left[\frac{2p}{1+p}\frac{C(\sqrt{\pi p})}{\sqrt{\pi p}} - \frac{1}{1+p}\frac{\sin(\pi p)}{\pi}\right], \tag{S22}$$

which limits:

$$\theta_{eq} = \begin{cases} \pi/4, & p = 0, \\ C(\sqrt{\pi})/\sqrt{\pi} \approx 34°, & p = 1. \end{cases} \tag{S23}$$

A useful symmetry exists for bends that start at 90°. Consider a partial Euler bend beginning at a tangent angle of 90° and spanning a total angle $\alpha$, described by the path $\theta'(l) = \pi/2 + \theta(l)$, where $\theta(l)$ is the propagation-angle function for the corresponding bend that begins at 0°. The equivalent propagation angle for this shifted path, $\theta_{eq,90}$, is related to the original equivalent angle, $\theta_{eq,0}$ through

$$\theta_{eq,90} = 45° - \theta_{eq,0}. \tag{S24}$$

This relationship is derived by replacing the path function $\theta(l)$ with a new path, $\theta'(l) = 90° + \theta(l)$, in the defining integrals of Eq. (S13). This relationship can be derived by substituting $\theta(l)$ in Eq. (S13) into $90° + \theta(l)$.

13

## 12. THERMO-OPTIC RESPONSE IN $SiO_2$-CLADDED DEVICE

The experimental results presented in this work were obtained using air-clad devices. To verify that the observed anisotropic TO behavior is not specific to air-cladded structures, we additionally evaluated the case of an $SiO_2$-cladded waveguide. The impact of replacing the air cladding with $SiO_2$ can be understood by examining the resulting changes in modal confinement. Because the refractive index contrast between the LN core and the $SiO_2$ cladding is lower than that between LN and air, the optical mode expands further into the surrounding oxide region. This redistribution produces two coupled effects: (1) Reduced confinement in the LN core, lowering both $\Gamma_e$ and $\Gamma_o$; (2) Increased overlap with oxide, increasing the overall $SiO_2$ (now representing both the buried oxide and top clad). These trends are confirmed by the simulation results in Figs. S16(a) and (b), which compare the confinement factors for air- and $SiO_2$-cladded structures across all propagation angles. The values were obtained using our analytical model (Eq. (3) in the main article), where the overlap factors $\Lambda_k$ were derived from the mode profile at $\theta = 0°$.

To quantify how these modified confinement factors influence TO efficiency, we recalculated the efficiencies for the $SiO_2$-cladded case using the same experimental fitting parameters (Fig. 4(e) in the main article). The predicted efficiencies for both cladding conditions are plotted in Fig. S16(c), normalized to the air-cladded $TE_0$ efficiency at $\theta = 0°$.

Although slight changes occur in the absolute efficiency values, the fundamental anisotropic behavior (dependence on the polarization and propagation angle) remains essentially unchanged. This is further supported by Fig. S16(d), where the percent difference between the two cases is found to be below 1% for the $TE_0$ mode and below 5% for the $TM_0$ mode. These small deviations confirm that the anisotropic TO trends identified in this study are robust and extend to both air- and $SiO_2$-cladded device architectures.

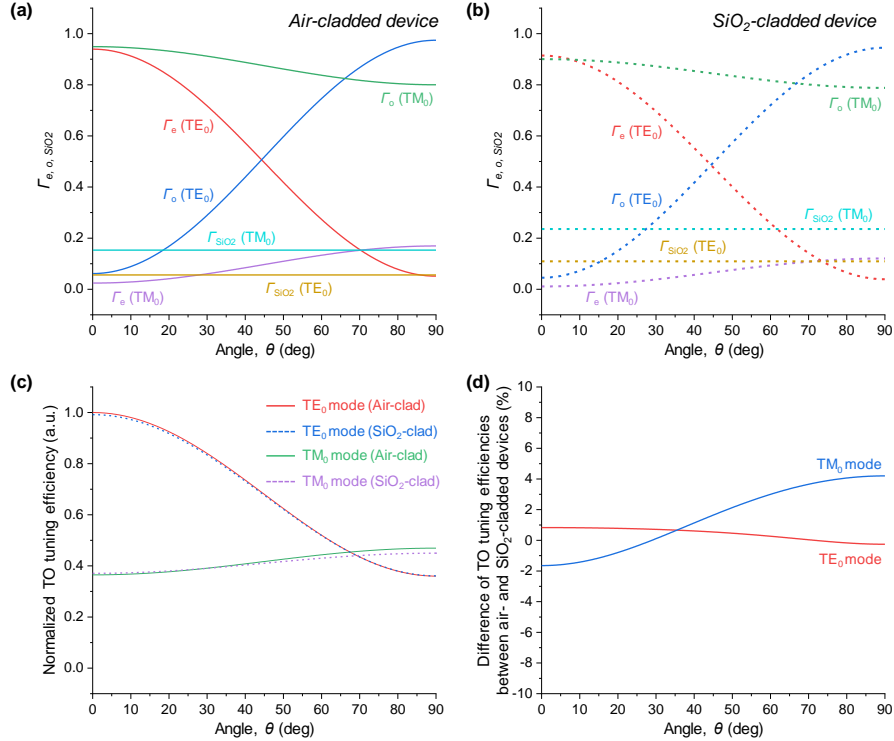

**Fig. S16.** Impact of adding an $SiO_2$ top cladding. (a, b) Confinement factors for the $TE_0$ and $TM_0$ modes, respectively, comparing air- and $SiO_2$-cladded structure. (c) Predicted TO tuning efficiencies for both cases, normalized to the air-cladded $TE_0$ efficiency at $\theta = 0°$. (d) Percent difference between the two cladding scenarios, showing $< 1\%$ deviation for the $TE_0$ and $< 5\%$ for $TM_0$.

## 13. $TM_0$-$TE_1$ MODE CONVERSION FOR ENHANCED THERMO-OPTIC EFFICIENCY

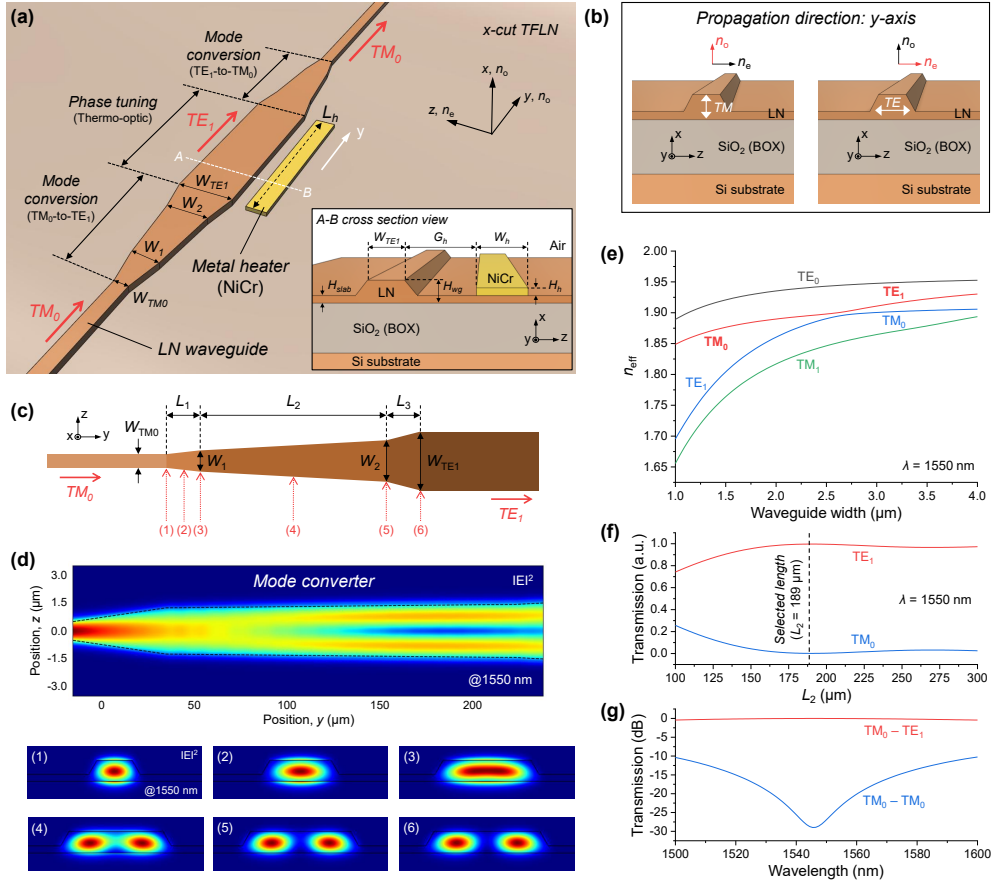

**Fig. S17.** Design and performance of the $TM_0$-to-$TE_1$ adiabatic mode converter. (a) Concept of the proposed phase shifter architecture, where an input $TM_0$ mode is converted to $TE_1$, thermally tuned, and converted back to $TM_0$. (b) TE and TM modes propagating along the $y$-axis, illustrating how the transverse field components project onto the ordinary and extraordinary axes. (c) Geometry of the three-section adiabatic taper with key lengths ($L_1$, $L_2$, $L_3$) and widths ($W_{TM_0}$, $W_1$, $W_2$, $W_{TE_1}$). (d) Simulated electric-field profiles $|\mathbf{E}|^2$ along the converter. Each cross-section mode profile, obtained using a finite difference eigenmode (FDE) solver (Lumerical), is shown depending on the position indicated in (c). (e) Effective indices of the three lowest-order eigenmodes versus waveguide width, showing mode hybridization. (f) Optimization of the central taper length ($L_2$). (g) Simulated spectral response of the optimized mode converter, achieving 99.6% conversion efficiency at 1550 nm and a 95%-efficiency bandwidth of 67 nm.

Through the systematic study in this work, we found that the TE mode propagating along the crystal $y$-axis exhibits significantly higher TO tuning efficiency than the TM mode. TO mitigate the inherently weaker TM response, we propose a mode-conversion-assisted architecture that temporarily converts the input $TM_0$ mode into the higher-order $TE_1$ mode, which possesses much stronger extraordinary-axis field overlap and therefore higher TO efficiency. As illustrated in Figs. S17(a) and S17(b), the light is first converted from $TM_0$ to $TE_1$, undergoes TO phase tuning, and is then reconverted back to $TM_0$ at the output. This strategy allows TM-polarized systems to benefit from the favorable TO properties of TE-like modes while remaining TM-compatible at the ports.

Below, we detail the design and optimization of the $TM_0$-to-$TE_1$ converter, followed by the confinement factor analysis demonstrating its performance advantage.

15

### 13-1. Mode converter design process

The converter exploits the natural mode hybridization between $TM_0$ and $TE_1$. To identify the required geometry, we first computed the effective indices of the lowest-order eigenmodes as a function of waveguide width (Fig. S17(e)). A strong hybridization occurs near a width of approximately 2.65 µm.

Using this hybridization regime, we designed a three-section adiabatic taper with lengths $L_1$, $L_2$ and $L_3$ (Fig. S17(c)). The first section ($L_1$) gradually expands the $TM_0$ input waveguide toward the hybridization region. The second section ($L_2$) facilitates the adiabatic $TM_0$-$TE_1$ mode conversion. The final section ($L_3$) transitions the waveguide to the output width required to stably support the $TE_1$ mode.

The key widths defining the taper profile are $W_{TM_0} = 1.0$ µm, $W_1 = 2.48$ µm, $W_2 = 2.85$ µm, and $W_{TE_1} = 3.0$ µm. The outer sections were fixed at $L_1 = 50$ µm and $L_3 = 15$ µm. To maximize conversion efficiency, the central taper length $L_2$ was optimized using an Eigenmode Expansion (EME) solver (Lumerical). As shown in Fig. S17(f), the optimal length is $L_2 = 189$ µm.

The spectral response of the optimized converter is shown in Fig. S17(g). At 1550 nm, the converter achieves a simulated conversion efficiency of 99.6%. The operational bandwidth, defined as the spectral range in which the efficiency exceeds 95%, is 67 nm.

### 13-2. Relative TO efficiency of the $TE_1$ mode converted from $TM_0$

The TO tuning efficiency of the $TM_0$ mode is less than half that of the $TE_0$ when propagating along the crystal $y$-axis (main text, Fig. 4(e)), substantially limiting the usefulness of TM-polarized devices. To quantify the benefit of the proposed conversion scheme, we calculated the confinement factors of the $TE_1$ mode in the widened phase-shifter waveguide ($W_{TE_1} = 3.0$ µm) and compared them with those of the $TM_0$ mode in its nominal geometry ($W_{TM_0} = 1.0$ µm).

The results are summarized in Table S1. The $TE_1$ mode exhibits a drastically stronger overlap with the extraordinary index region ($\Gamma_e = 0.924$) and correspondingly weaker overlap with the ordinary index region ($\Gamma_o = 0.052$). This distribution closely resembles that of the highly efficient $TE_0$ mode and stands in significant contrast to the $TM_0$ mode, which is dominated by ordinary-axis overlap.

Substituting these confinement factors into Eq. (1) of the main article shows that the $TE_1$ mode achieves a normalized TO tuning efficiency of approximately 99% relative to the $TE_0$ mode. This demonstrated that the proposed mode-conversion architecture effectively enables nearly $TE_0$-level performance, while preserving $TM_0$ input/output compatibility in the photonic circuit.

| Confinement Factor | $TM_0$ ($W_{TM_0} = 1.0$ µm) | $TE_1$ ($W_{TE_1} = 3.0$ µm) |
|---|---|---|
| $\Gamma_e$ | 0.024 | 0.924 |
| $\Gamma_o$ | 0.945 | 0.052 |
| $\Gamma_{BOX}$ | 0.150 | 0.089 |

**Table S1.** Calculated optical confinement factors for the $TM_0$ and $TE_1$ modes. The factors correspond to the regions with extraordinary ($\Gamma_e$) and ordinary ($\Gamma_o$) refractive indices of the TFLN core, and the buried oxide layer ($\Gamma_{BOX}$).



\end{document}